\documentclass[prd,onecolumn,showpacs,superscriptaddress,nofootinbib,floatfix,11pt]{revtex4-2}

\usepackage[utf8]{inputenc}
\usepackage{amsmath,amssymb}
\usepackage{slashed}
\usepackage[colorlinks=true, 
linkcolor=blue,
breaklinks=true,
urlcolor=magenta,
citecolor=blue]{hyperref}
\usepackage[usenames,dvipsnames]{color}
\usepackage{appendix}
\usepackage{braket,bm}
\usepackage{multirow}
\usepackage{enumitem}
\usepackage{color}
\usepackage{cancel}
\usepackage{orcidlink}
\usepackage{mathrsfs}
\usepackage{graphicx}
\usepackage{tikz}
\usepackage[normalem]{ulem}
\usepackage{booktabs}
\usepackage{array}
\usepackage{overpic}
\usepackage{float}
\usepackage{threeparttable}
\allowdisplaybreaks[4]
\graphicspath{{figs/}}

\usetikzlibrary{calc}
\usetikzlibrary{intersections}
\usetikzlibrary{trees}
\usetikzlibrary{decorations.pathmorphing}
\usetikzlibrary{decorations.markings}
\usetikzlibrary{arrows.meta}
\usetikzlibrary{patterns}
\tikzset{
   global scale/.style={
      scale=#1,
      every node/.append style={scale=#1}},
   photon/.style={
   decorate,decoration={
      snake,
      amplitude=0.5mm,
      segment length=2.5mm,post length=0,pre length=0},draw=red},
   nucleon/.style={draw=black, postaction={decorate},
      decoration={markings,mark=at position .65 with{\arrow[draw=black]{latex}}}},
   pion/.style={draw=blue, postaction={decorate},
      decoration={markings,mark=at position .55 with{\arrow[draw=blue]{}}}},
    nucleonstar/.style={draw=black, postaction={decorate},
      decoration={markings, mark=at position 0.7 with {\arrow[draw=black]{latex}}}},
    }

\newcommand{\itp}{\affiliation{ Institute of Theoretical Physics, Chinese Academy of Sciences, Beijing 100190, China}}

\newcommand{\ucas}{\affiliation{School of Physical Sciences, University of Chinese Academy of Sciences, Beijing 100049, China}}

\newcommand{\julich}{\affiliation{Forschungszentrum J\"ulich, Institute for Advanced Simulation, 52425 J\"ulich, Germany}}

\newcommand{\bonn}{\affiliation{Helmholtz-Institut f\"ur Strahlen- und Kernphysik (Theorie) and\\
Bethe Center for Theoretical Physics, Universit\"at Bonn, 53115 Bonn, Germany}}

\newcommand{\md}{\mathrm{d}}
\newcommand{\mF}{\mathcal{F}}

\begin{document}
\title{\boldmath{Dispersive analysis of the $J/\psi\to\pi^0 \gamma^\ast$ transition form factor\\ with $\rho$--$\omega$ mixing effects}}

\author{Xiong-Hui Cao\orcidlink{0000-0003-1365-7178}}
\email{xhcao@itp.ac.cn}
\itp

\author{Feng-Kun Guo\orcidlink{0000-0002-2919-2064}}\email{fkguo@itp.ac.cn}
\itp\ucas

\author{Christoph Hanhart\orcidlink{0000-0002-3509-2473}}\email{c.hanhart@fz-juelich.de}
\julich

\author{Bastian Kubis\orcidlink{0000-0002-1541-6581}}\email{kubis@hiskp.uni-bonn.de}
\bonn


\begin{abstract}
Motivated by the discrepancies noted recently between the theoretical predictions of the electromagnetic $J/\psi \to \pi^0 \gamma^*$ transition form factor and the BESIII data, we reanalyze this transition form factor using the dispersive Khuri--Treiman equations, with final-state interactions in both the direct channel and the crossed channels properly considered. 
This improved framework incorporates $\rho$--$\omega$ mixing effects. The effect of four-pion states is evaluated through a dispersively improved vector-meson-dominance model.
From this information, we propose a two-parameter fit that provides an excellent description of the BESIII data over the broad energy range from 0 to 2.8~GeV. 
We demonstrate that the $\rho\pi^0$ decay mode of the $J/\psi$ is dominated by strong interaction, while the $\omega\pi^0$ mode is dominated by one-photon exchange. From this, we extract the relative phase between the strong and the one-virtual-photon (electromagnetic) modes in hadronic decays of $J/\psi$ as $(62 \pm 21)^{\circ}$.  This could provide useful information in understanding the long-standing $\rho \pi$ puzzle in $J/\psi$ decays.
\end{abstract}

\maketitle
\tableofcontents

\section{Introduction}

Radiative decays of the charmonium state $J/\psi$ serve as a crucial probe for investigating nonperturbative dynamics in quantum chromodynamics (QCD). Since the $J/\psi$ lies more than 600~MeV below the threshold for Okubo--Zweig--Iizuka (OZI) allowed open-charm channels, its total decay width is rather small---less than 100~keV---which means that its radiative decays to light hadrons constitute significant branching fractions that can be measured with good precision.
In particular, the branching fraction for inclusive radiative $J/\psi$ decays is $(8.8 \pm 1.1) \%$~\cite{ParticleDataGroup:2024cfk,CLEO:2008gct}.
In addition to the production of light hadrons and real photons, as in the $J/\psi \to P\gamma$ process where $P=\pi^0, \eta^{(\prime)}$, the BESIII experiment also allows for measurements of the corresponding electromagnetic Dalitz decay processes~\cite{BESIII:2014dax}, where a time-like virtual photon is emitted, producing a detectable $e^{+} e^{-}$ pair, i.e., $J / \psi \to Pe^{+} e^{-}$. 
This differential decay width measures the transition form factor (TFF) $f_{\psi \pi^0}$, which encodes the nonperturbative strong interaction dynamics governing the $J/\psi\to\pi^0\gamma^*$ transition.

The interactions of hadrons with both real and virtual photons are often well-described within the framework of vector meson dominance (VMD), where the photon couples to hadrons through intermediate vector mesons~\cite{Sakurai:1960ju}. 
The TFF for $J/\psi \to P\gamma^*$ is sometimes described using a simple monopole parameterization, with the pole corresponding to a characteristic charmonium mass.
Although the charmonium mass scale for the monopole form factor should only be valid in the large momentum-transfer region, this form was assumed in Ref.~\cite{Fu:2011yy} for all three pseudoscalar ($\pi^0$, $\eta$, and $\eta'$) final states, and the corresponding branching fractions were estimated.
Interestingly, for $\eta$ and $\eta'$, the experimental results agree well with these predictions. 
This is because the emitted photon mainly originates from the initial charm quark, rather than from light quarks that would need to be converted from at least three gluons or one photon. 
That the latter is highly suppressed can be seen from the extremely narrow width of the $J/\psi$. 
Emitting a photon from the charm or anticharm quark allows the $J/\psi$ to be converted into a virtual pseudoscalar intermediate state of $c\bar{c}$, which can then annihilate into an isoscalar pseudoscalar meson via two gluons. 
For the importance of the intermediate $\eta_c$ state in $J/\psi \to \eta^{(\prime)}\gamma^*$, see phenomenological analyses in Refs.~\cite{Zhao:2010mm,Chen:2014yta,Yan:2023nqz}. 
For lattice QCD calculations of the $J/\psi\to\eta^{(\prime)}\gamma^{*}$ TFFs, we refer to Refs.~\cite{Jiang:2022gnd,Shi:2024fyv,Batelaan:2025vbb,Batelaan:2025vhx}. 

In contrast to $J/\psi \to \eta^{(\prime)}\gamma^*$, the $J/\psi \to \pi^0 \gamma^*$ process should be dominated by a mechanism where the photon is emitted from light quarks, since otherwise the production of the isovector pion from gluons would require isospin symmetry breaking.
In this case, the TFF is dominated by light-quark degrees of freedom, and the simple monopole parameterization with a charmonium pole is inappropriate for describing the $J/\psi \to \pi^0 \gamma^*$ TFF.
Isospin-breaking effects can still play a role, particularly in the region around the $\omega$ resonance, where it has been well-established that $\rho$--$\omega$ mixing can lead to a distortion of the pion electromagnetic form factor as observed in $e^{+} e^{-}$ annihilations and in $\eta'\to \gamma\pi^+\pi^-$~\cite{Holz:2022hwz}.
Such mixing can play a role in the most significant hadronic vacuum polarization contribution to the muon anomalous magnetic moment $(g-2)_\mu$~\cite{Aliberti:2025beg}.
Recently, the BESIII Collaboration~\cite{BESIII:2025xjh} reported the first measurement of the TFF and an improved measurement of the branching fraction of this electromagnetic Dalitz decay $J / \psi \to e^{+} e^{-} \pi^0$ in the full $m_{e^{+} e^{-}}$ spectrum using a data sample of $10^{10}~J / \psi$ events. 
A clear $\rho$--$\omega$ interference structure has been identified.
Consequently, there is an urgent need for a consistent and model-independent analysis of the $J / \psi \pi^0$ TFF, taking into account the effects of $\rho$--$\omega$ mixing. This work extends the analysis in Ref.~\cite{Kubis:2014gka} accordingly.
The dispersive Khuri--Treiman (KT) equations~\cite{Khuri:1960zz} for $J / \psi \to 3 \pi$ are utilized, with final-state interactions in both the direct channel and crossed channels properly considered. Various contributions of the $2 \pi$, $3 \pi$, $4 \pi$, and $c \bar{c}$ intermediate states will be systematically accounted for.

The paper is organized as follows. In Sec.~\ref{sec:Def}, we briefly discuss the kinematics and introduce the $J/\psi\to \pi^0 \gamma^*$ TFF. In Sec.~\ref{sec:Disp}, we first review the KT formalism for the $J / \psi \to 3 \pi$ decay and show its connection to the $J/\psi\to \pi^0 \gamma^*$ TFF. 
We then develop a consistent representation incorporating $\rho$--$\omega$ mixing effects, followed by estimates of both the $4\pi$ continuum and charmonium contributions. 
Section~\ref{sec:pQCD} examines the asymptotic behavior of the TFF and derives relevant sum rule. In Sec.~\ref{sec:Results}, the BESIII data are fitted, and our conclusions are given in Sec.~\ref{sec:Summary}. 
Appendix~\ref{app:dipole} provides additional details on the charmonium contribution through dipole form factor fits.

\section{Definitions and kinematics}\label{sec:Def}

We consider the decay of the vector meson $J/\psi$ into a $\pi^0$ and a dilepton pair (mediated by a virtual photon $\gamma^\ast$). The $J/\psi\to\pi^0 \gamma^\ast$ TFF is defined according to the matrix element
\begin{align}
    \left\langle\pi^0(p_0)\right| j_\mu(0)\left|\psi\left(p_V, \lambda\right)\right\rangle=-i \epsilon_{\mu \nu \alpha \beta} \epsilon^{\nu}\!\left(p_V, \lambda\right) p_0^\alpha q^\beta f_{\psi \pi^0}(s),
\end{align}
where $j_\mu=\sum_f Q_f \bar{q}_f \gamma_\mu q_f$ denotes the electromagnetic current, $Q_f$ is the charge of the quarks of different flavors ($f$), $\epsilon_{\mu \nu \alpha \beta}$ is the Levi-Civita tensor, $\epsilon^{\nu}\!\left(p_V, \lambda\right)$ is the polarization vector of the $J/\psi$ with helicity $\lambda$, $q=p_V-p_0$ with $s\equiv q^2>0$, and $f_{\psi \pi^0}(s)$ is the electromagnetic TFF of the $J/\psi \to \pi^0 \gamma^*$ transition. One also uses the corresponding normalized TFF
\begin{align}\label{eq:normalized TFF}
    F_{\psi \pi^0}(s)=\frac{f_{\psi \pi^0}(s)}{f_{\psi \pi^0}(0)}.
\end{align}
The $J/\psi\to \pi^0\ell^+ \ell^-$ ($\ell=e,\mu$) amplitude is given by~\cite{Landsberg:1985gaz}
\begin{align}\label{eq:Mpsipi}
    \mathcal{M}_{\psi \pi^0}=4\pi \,i\, \alpha \,\epsilon_{\mu \nu \alpha \beta} \epsilon^\mu(p_V,\lambda) p_0^\nu q^\alpha \frac{f_{\psi \pi^0}(s)}{s} \bar{u}_s\left(p_{\ell^{-}}\right) \gamma^\beta v_{s^{\prime}}\left(p_{\ell^{+}}\right),
\end{align}
where $q=p_{\ell^+}+p_{\ell^-}$, and the fine structure constant $\alpha=e^2/(4\pi)$. The differential decay rate in terms of the $J/\psi\to\pi^0 \gamma^\ast$ TFF, normalized to the partial width into $\pi^0$ and a real photon, can be written as
\begin{align}\label{eq:diff_Gamma}
  \frac{1}{\Gamma_{\psi \rightarrow \pi^0 \gamma} } \left( \frac{\mathrm{d} \Gamma_{\psi \rightarrow \pi^0 \ell^{+} \ell^{-}}}{\mathrm{~d} s}\right)=\frac{16 \alpha}{3 \pi}\left(1+\frac{2 m_{\ell}^2}{s}\right) \frac{q_{\ell}(s) q_{\psi \pi^0}^3(s)}{\big(M_\psi^2-M_{\pi^0}^2\big)^3}\left|F_{\psi \pi^0}(s)\right|^2 \, ,
\end{align}
where the masses of $J/\psi$, $\pi^0$, and the leptons $\ell$ are denoted by $M_\psi$, $M_{\pi^0}$, and $m_\ell$, respectively. The corresponding real photon width determines the normalization of the TFF as
\begin{align}\label{eq:pi0gamma_decay}
    \Gamma_{\psi \rightarrow \pi^0 \gamma}=\frac{\alpha\big(M_\psi^2-M_{\pi^0}^2\big)^3}{24 M_\psi^3}\left|f_{\psi \pi^0}(0)\right|^2.
\end{align}
Finally, the momenta are given by
\begin{align}
    q_{\ell}(s)=\frac{1}{2} \sqrt{s-4 m_{\ell}^2}, \quad q_{A B}(s)=\frac{\lambda^{1 / 2}\!\left(s, M_A^2, M_B^2\right)}{2 \sqrt{s}},
\end{align}
where $\lambda(a, b, c)=a^2+b^2+c^2-2(a b+a c+b c)$ is the standard K\"all\'en triangle function.

In formulating a dispersion relation for the $J/\psi \rightarrow \pi^0 \gamma^\ast$ TFF, the three-pion decay $J/\psi\left(p_V\right)\to \pi^0\left(p_0\right)\pi^{+}\left(p_{+}\right) \pi^{-}\left(p_{-}\right)$ plays a pivotal role. The helicity amplitude for the three-pion decay can be expressed in terms of a Lorentz-invariant scalar function $\mathcal{F}(s,t,u)$ that encodes the dynamics,
\begin{align}\label{eq:M3pi}
    \mathcal{M}_{\psi\to 3\pi}^\lambda(s, t, u)=i \epsilon_{\mu \nu \alpha \beta} \epsilon^\mu(p_V,
    \lambda) p_0^\nu p_{+}^\alpha p_{-}^\beta \mathcal{F}(s, t, u).
\end{align}
The Mandelstam variables are defined as 
\begin{align}
    \begin{aligned}
    s & =\left(p_{+}+p_{-}\right)^2=\left(p_V-p_0\right)^2, \\
    t & =\left(p_{-}+p_0\right)^2=\left(p_V-p_{+}\right)^2, \\
    u & =\left(p_{+}+p_0\right)^2=\left(p_V-p_{-}\right)^2,
    \end{aligned}
\end{align}
which satisfy the usual identity $s+t+u=M_\psi^2+3M_\pi^2\equiv 3s_0$. For the remainder of this manuscript, we will set $M_{\pi^0}=M_{\pi^\pm}$ unless otherwise specified for isospin breaking, for which we take the charged pion mass $M_\pi=M_{\pi^\pm}$ as reference. The $s$-, $t$-, and $u$-channel processes are identical in the limit of isospin symmetry. Therefore, we will now focus on the $s$-channel.

The scattering angle in the $s$-channel, defined in the center-of-mass (c.m.) frame of the $\pi^+ \pi^-$ pair, is denoted by $\theta_s$,
\begin{align}\label{eq:theta}
    \cos( \theta_s(s, t, u))=\frac{t-u}{4 q_{\pi\pi}(s) q_{\psi\pi}(s)}\equiv\frac{t-u}{\kappa(s)},\quad \sin \theta_s=\frac{\sqrt{\phi(s, t, u)}}{2 \sqrt{s} q_{\pi\pi}(s) q_{\psi\pi}(s)}.
\end{align}
The zeros of the well-known Kibble function~\cite{Kibble:1960zz},
\begin{align}
    \phi(s, t, u)=\left(2 \sqrt{s} \sin \theta_s q_{\pi\pi}(s) q_{\psi\pi}(s)\right)^2=s t u-M_\pi^2\left(M_{\psi}^2-M_\pi^2\right)^2,
\end{align}
determine the boundaries for $s$ in the physical decay region of the
process by solving $\phi(s, t, u)=0$,
\begin{align}
    s_{\min }=4 M_\pi^2, \quad s_{\max }=\left(M_\psi-M_\pi\right)^2,
\end{align}
while the Dalitz-plot boundaries of $t(s)$ for a given value of $s$ can be obtained from Eq.~\eqref{eq:theta} by taking $\cos\theta_s=\pm 1$,
\begin{align}
    t_{\max , \min }(s)=\frac{M_{\psi}^2+3 M_\pi^2-s}{2} \pm 2 q_{\pi\pi}(s) q_{\psi\pi}(s).
\end{align}
The $s$-channel partial-wave decomposition of the helicity amplitude $\mathcal{M}_{\psi\to 3\pi}^\lambda(s, t, u)$ is given by~\cite{Jacob:1959at}
\begin{align}
    \mathcal{M}_{\psi\to 3\pi}^\lambda(s, t, u)=\sum^\infty_{J~\text{odd}}(2J+1)d^J_{\lambda 0}(\theta_s)h_J^{\lambda}(s),
\end{align}
where $d_{\lambda 0}^J\left(\theta_s\right)$ are the Wigner $d$-functions. Due to Bose symmetry, the sum over partial waves is restricted to odd values of $J$, and parity conservation implies that $h^0_J(s)=0$ and $h^{+1}_J(s)=-h^{-1}_J(s) \equiv h_J(s)$. Therefore, there is only one independent helicity amplitude. One can rewrite the partial-wave expansion for the invariant amplitude $\mathcal{F}(s,t,u)$ in the following form:
\begin{align}\label{eq:PW}
    \mathcal{F}(s, t, u)=\sum^\infty_{J~\text{odd}}(q_{\pi\pi}(s)q_{\psi\pi}(s))^{J-1} P_J^{\prime}\!\left(z_s\right) f_J(s),
\end{align}
where $z_s=\cos\theta_s$, and the kinematic-singularity-free amplitudes $f_J(s)$ are related to $h_J(s)$ as
\begin{align}
    f_J(s) \equiv \sqrt{\frac{2}{s}} \frac{2 J+1}{\sqrt{J(J+1)}} \frac{h_J(s)}{(q_{\pi\pi}(s)q_{\psi\pi}(s))^J}.
\end{align}
Finally, the measured differential decay width can be calculated in terms of Eq.~\eqref{eq:M3pi} by
\begin{align}\label{eq:JpsiTo3pi_width}
    \frac{\md^2 \Gamma_{\psi\to3\pi}}{\md s \md t}(s, t)&= \frac{1}{(2 \pi)^3} \frac{1}{32 M_\psi^3}\frac{1}{3}\sum_\lambda|\mathcal{M}^\lambda_{\psi\to 3\pi}(s, t)|^2\nonumber\\
    &= \frac{1}{(2 \pi)^3} \frac{1}{32 M_{\psi}^3} \frac{1}{3} \frac{\phi(s, t, u)}{4}|\mathcal{F}(s, t, u)|^2 .
\end{align}

\section{Dispersive formalism for the
transition form factor}\label{sec:Disp}

Dispersion relations provide a framework for constructing form factors from their discontinuities across the physical cut along the positive real $s$ axis; see Refs.~\cite{Oller:2019rej, Yao:2020bxx} for recent reviews. For the $J/\psi\to\pi \gamma^\ast$ TFF, an unsubtracted dispersion relation can, in principle, be employed~\cite{Koepp:1974da},
\begin{align}\label{eq:0DR}
    f_{\psi\pi^0}(s) = \frac{1}{2\pi i} \int_{4M_\pi^2}^\infty \md s^\prime~ \frac{\operatorname{disc} f_{\psi\pi^0}(s^\prime)}{s^\prime-s-i\epsilon},
\end{align}
where contributions to the discontinuity are given by multiparticle intermediate states as well as possible single-particle pole contributions. 
The convergence of Eq.~\eqref{eq:0DR} is ensured by the constituent counting rule~\cite{Matveev:1973ra,Brodsky:1973kr,Brodsky:1974vy} and confirmed by perturbative QCD (pQCD) for large momentum transfers~\cite{Lepage:1979zb,Lepage:1980fj}. 
The scaling behavior of the form factors at large $s$ can be easily understood from the scaling behavior of the cross section $\sigma\left(e^{+} e^{-} \rightarrow J/\psi \pi^0\right)$~\cite{Fang:2021wes} [cf.\ Eq.~\eqref{eq:Mpsipi}]
\begin{align}\label{eq:CCR}
    \sigma\left(e^{+} e^{-} \rightarrow J / \psi \pi^0\right) &=\frac{\pi \alpha^2}{6 s^3} \lambda^{\frac{3}{2}}\left(s, M_\psi^2, M_\pi^2\right)\left|f_{\psi \pi}(s)\right|^2 \nonumber\\
    &\propto\frac{1}{s^4},\quad \text{when}\ s\to \infty.
\end{align}
In particular, the $1/s^4$ scaling arises from the necessity of a helicity flip since the vector meson can only be transversely polarized. Note that our analysis is confined to low-energy and excludes high-energy effective VMD-type diagrams due to one more virtual-photon exchange; such diagrams would contribute a $1/s^2$ scaling~\cite{Guo:2016fqg} instead of the $1/s^4$ behavior considered here. Utilizing Eq.~\eqref{eq:CCR}, the form factor $f_{\psi\pi^0}(s)$ must fall off as $1/s^2$.

Setting $s=0$ in Eq.~\eqref{eq:0DR} leads to a sum rule for the normalization $f_{\psi\pi^0}(0)$,
\begin{align}\label{eq:sum rule_v1}
    f_{\psi\pi^0}(0) = \frac{1}{2\pi i} \int_{4M_\pi^2}^\infty \md s^\prime~ \frac{\operatorname{disc} f_{\psi\pi^0}(s^\prime)}{s^\prime}.
\end{align}
Alternatively, one can use a once-subtracted dispersion relation to reduce the sensitivity to high-energy contributions in the dispersive integral,
\begin{align}\label{eq:1DR}
    f_{\psi\pi^0}(s) = f_{\psi\pi^0}(0)+\frac{s}{2\pi i} \int_{4M_\pi^2}^\infty \md s^\prime~ \frac{\operatorname{disc} f_{\psi\pi^0}(s^\prime)}{s^\prime(s^\prime-s-i\epsilon)}.
\end{align}
If the subtraction constant $f_{\psi\pi^0}(0)$ in Eq.~\eqref{eq:1DR} (which is in principle arbitrary) satisfies the sum rule in Eq.~\eqref{eq:sum rule_v1}, then Eq.~\eqref{eq:1DR} becomes fully equivalent to Eq.~\eqref{eq:0DR}.
Equations~\eqref{eq:0DR} and~\eqref{eq:1DR} serve as starting points for the following discussion, mainly focusing on calculating the discontinuity of the TFF from the lightest multipion intermediate states, $2\pi$, $3\pi$ (approximated by $\omega$, $\phi$),\footnote{The $\phi$ couples dominantly to $K \bar{K}$, but since the $\phi$ has a width of only about 4~MeV, we will treat its width as a constant and neglect the analytic structure due to the $K\bar K$ intermediate states. For a discussion of the role of $K \bar{K}$ contributions in the $J/\psi \to 3 \pi$ process, see Ref.~\cite{Guo:2010gx}.} $4\pi$ [approximated by $\rho^\prime(1450)$], and charmonium(-like) states. We will elaborate on these in the following sections.

\subsection{Two-pion intermediate states: Khuri--Treiman representation}

\begin{figure}[t]
    \centering
    \includegraphics[width=.7\linewidth]{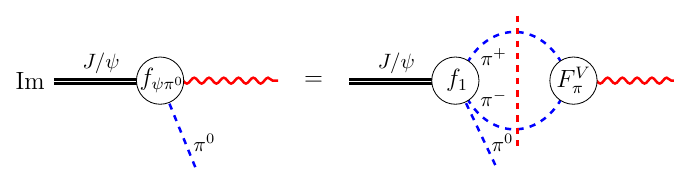}
    \caption{Unitarity relation for the $J/\psi\to\pi^0\gamma^\ast$ TFF $f_{\psi\pi^0}$. The blue dashed lines denote pions, the wiggly lines represent photons, the double-solid lines represent the $J/\psi$, and the red dashed line indicates that the intermediate $\pi\pi$ states are to be taken on shell.}\label{fig:2piDisc}
\end{figure}

The contribution of the two-pion intermediate states to the discontinuity of the $J/\psi\to \pi^0 \gamma^\ast$ TFF, shown in Fig.~\ref{fig:2piDisc}, is given by~\cite{Koepp:1974da}
\begin{align}\label{eq:2piDisc}
    \frac{1}{2i}\operatorname{disc}f_{\psi\pi^0}^{(2\pi)}(s)=\frac{s \sigma_{\pi}^3(s)}{96\pi}F^{V *}_\pi(s) f_1(s),
\end{align}
where $\sigma_\pi(s)=2q_{\pi\pi}(s)/\sqrt{s}$ is the two-pion phase space factor, and $F_\pi^V(s)$ is the pion vector form factor. The analytical solution for $F_\pi^V(s)$ is given in terms of the Omn\`es function~\cite{Omnes:1958hv},
\begin{align}\label{eq:Omnes}
    F_\pi^V(s) & =P(s) \Omega(s),\quad\Omega(s) =\exp \left\{\frac{s}{\pi} \int_{4 M_\pi^2}^{\infty} \md s^{\prime} \frac{\delta\!\left(s^{\prime}\right)}{s^{\prime}\!\left(s^{\prime}-s\right)}\right\},
\end{align}
with a real-valued subtraction polynomial $P(s)$. The pion vector form factor is expected to behave as $F_\pi^V(s) \asymp 1/s$ at large energies, as suggested by pQCD (up to logarithmic corrections), and to be free of zeros~\cite{Leutwyler:2002hm,Ananthanarayan:2011xt}. Thus, $P(s)$ is a constant and can be set to 1 due to gauge invariance ($F_\pi^V(0)=1$). 
Note that for a full analysis of the pion vector form factor, one needs to account for $\rho$--$\omega$ mixing, which will be discussed in the next section.

\begin{figure}[t]
    \centering
    \includegraphics[width=.5\linewidth]{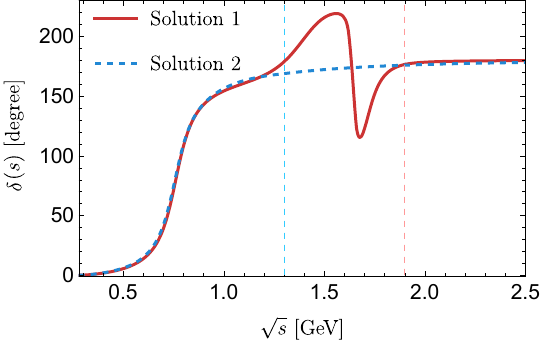}
    \caption{Solution~1 for the $P$-wave phase input $\delta(s)$ from Ref.~\cite{Schneider:2012ez}, extracted from the pion vector form factor and valid up to roughly 1.9~GeV. Solution~2~\cite{Pelaez:2024uav} is the elastic $\pi\pi$ $P$-wave phase shift and is valid up to roughly 1.3~GeV; see the main text for further details.}
    \label{fig:phase}
\end{figure}

From Eq.~\eqref{eq:Omnes}, we observe that the Omn\`es function is completely determined by the phase input. To estimate the associated uncertainties, especially above the low-energy elastic region, we compare two different parameterizations, referred to as ``Solution~1"~\cite{Schneider:2012ez} and ``Solution~2"~\cite{Pelaez:2024uav}, as shown in Fig.~\ref{fig:phase}. 
Solution~1 employs the phenomenological parameterization of the pion vector form factor proposed in Ref.~\cite{Roig:2011iv} (see also Ref.~\cite{Gonzalez-Solis:2019iod}, as well as Ref.~\cite{RuizArriola:2024gwb} for a direct extraction of the form factor phase). 
In Ref.~\cite{Schneider:2012ez}, the form factor was constrained by fitting the experimental data to extract the corresponding phase, which was then smoothly matched to the Roy equation phase-shift solution below 1~GeV; accordingly, the $\rho^\prime$ and $\rho^{\prime\prime}$ region is already encoded effectively in $\Omega(s)$ within an elastic approximation. Solution~2 utilizes the latest phase shifts derived from the dispersive analysis of $\pi\pi$ scattering. 
Given that the $P$-wave inelasticity in Ref.~\cite{Pelaez:2024uav} is found to be rather small up to approximately 1.3~GeV, we refrain from considering it. For our analysis, beyond $\Lambda_\delta=1.9$~GeV for solution~1 and $\Lambda_\delta=1.3$~GeV for solution~2, we smoothly guide $\delta$ to the asymptotic value $\pi$ using~\cite{Moussallam:1999aq,Gonzalez-Solis:2019iod}
\begin{align}
    \delta_{\infty}(s) \equiv \lim _{s \rightarrow \infty} \delta(s)=\pi-\frac{\alpha}{\beta+\left(s / \Lambda_\delta^2\right)^{3/2}},
\end{align}
where $\alpha$ and $\beta$ are parameters introduced to ensure that both the phase $\delta(s)$ and its first derivative $\delta^{\prime}(s)$ are continuous at $s=\Lambda_\delta^2$. 
Their explicit expressions are
\begin{align}
    \alpha=\frac{3\left(\pi-\delta\!\left(\Lambda_\delta^2\right)\right)^2}{2 \Lambda_\delta^2 \delta^{\prime}\!\left(\Lambda_\delta^2\right)}, \quad \beta=-1+\frac{3\left(\pi-\delta\!\left(\Lambda_\delta^2\right)\right)}{2 \Lambda_\delta^2 \delta^{\prime}\!\left(\Lambda_\delta^2\right)} .
\end{align}
We choose solution~1 as the input for the central value and use solution~2 for error estimation.
It is evident from Fig.~\ref{fig:phase} that these two solutions of $\delta$ lead to a noticeable difference only in the higher resonance region, specifically around the $\rho^\prime(1450)$ and $\rho^{\prime\prime}(1700)$ resonances. 
In particular, the vector radius of the pion is well-reproduced from Eq.~\eqref{eq:Omnes}~\cite{Guo:2008nc},
\begin{align}
    \left\langle\left(r_\pi^V\right)^2\right\rangle=\left.6 \frac{\md F_\pi^V(s)}{\md s}\right|_{s=0}=\frac{6}{\pi} \int_{4 M_\pi^2}^{\infty} \md s \frac{\delta\!\left(s\right)}{s^{2}}=0.419(1)~\mathrm{fm}^2,
\end{align}
which differs from the comprehensive dispersive result $\left\langle\left(r_\pi^V\right)^2\right\rangle=0.429(4)~\mathrm{fm}^2$~\cite{Colangelo:2018mtw} by about 2\%, where the minor discrepancy comes from other inelastic effects ($4 \pi$, $\pi \omega$, $K\bar{K}$, \ldots) in the weighted integral of the $P$-wave phase shift. 

Besides the pion vector form factor we also need the $P$-wave kinematic-singularity-free $J/\psi\to 3\pi$ decay amplitude $f_1$ (see Fig.~\ref{fig:2piDisc}). It has previously been studied within the context of the Veneziano model~\cite{Szczepaniak:2014bsa} and the $N/D$ method~\cite{Guo:2010gx}. 
The analytical structure of this amplitude has been investigated using dispersion relations in Refs.~\cite{Kubis:2014gka, JPAC:2023nhq}. Here, we adopt the KT framework, which is briefly reviewed for $f_1$ below.

According to Eq.~\eqref{eq:PW}, each term in the sum is a polynomial in the variables $t$ and $u$. The singularities of $\mathcal{F}(s,t,u)$ in these variables, required by $t$- and $u$-channel unitarity, can arise only from an infinite number of terms in the sum. 
In practice, however, it is necessary to truncate the sum to a finite number of partial waves at low energies. Such a truncation introduces a violation of the crossing symmetry or, in other words, breaks the analytic properties of the crossed $t$- and $u$-channels. 
This issue can be partially resolved by employing the KT equation formalism,\footnote{Another way to restore crossing symmetry is the Roy(-like) equations~\cite{Roy:1971tc}, whose applications have been extended to analyze lattice QCD data at unphysical pion masses in Refs.~\cite{Cao:2023ntr,Rodas:2023nec,Cao:2024zuy,Cao:2025hqm}. Interestingly, for $\pi\pi$ scattering, when both formalisms are restricted to only $S$- and $P$-waves, the KT equation has been shown to be formally equivalent to the Roy equation~\cite{Albaladejo:2018gif}.} which replaces the infinite sum of partial waves in the $s$-channel with three truncated sums of single-Mandelstam-variable amplitudes analogous to Eq.~\eqref{eq:PW}, resulting in
\begin{align}
    \mathcal{F}(s, t, u)=&\sum^{J_\text{max}}_{J~\text{odd}}(q_{\pi\pi}(s)q_{\psi\pi}(s))^{J-1} P_J^{\prime}\!\left(z_s\right) F_J(s) \nonumber\\
    &+\sum^{J_\text{max}}_{J~\text{odd}}(q_{\pi\pi}(t)q_{\psi\pi}(t))^{J-1} P_J^{\prime}\!\left(z_t\right) F_J(t)+\sum^{J_\text{max}}_{J~\text{odd}}(q_{\pi\pi}(u)q_{\psi\pi}(u))^{J-1} P_J^{\prime}\!\left(z_u\right) F_J(u),
\end{align}
where $z_t=(s-u)/\kappa(t)$ and $z_u=(t-s)/\kappa(u)$.

By truncating each sum to $J_\text{max} = 1$, we obtain the simplest crossing-symmetric KT decomposition~\cite{Niecknig:2012sj,Hoferichter:2012pm,Danilkin:2014cra}:
\begin{align}\label{eq:KT_decomposition}
    \mathcal{F}(s, t, u)=\mathcal{F}(s)+\mathcal{F}(t)+\mathcal{F}(u),
\end{align}
where the simplified notation $\mathcal{F}\equiv \mathcal{F}_1$ is used. Each single-variable amplitude has only the right-hand cut in its respective Mandelstam variable. 
These decompositions are also known as reconstruction theorems, justified in chiral perturbation theory at a given order using fixed-variable dispersion relations~\cite{Stern:1993rg, Knecht:1995tr, Bijnens:2007pr, Zdrahal:2008bd, Stamen:2022eda}. 
Note that $\mathcal{F}(s)$ and $f_1(s)$ share the same discontinuities along the right-hand cut, $\operatorname{disc}\mathcal{F}(s)=\operatorname{disc}f_1(s)$. We can rewrite the elastic unitarity relation of $f_1(s)$~\cite{Niecknig:2012sj},
\begin{align}
    \operatorname{disc} f_1(s)=2 i f_1(s) \sin \delta(s)\, e^{-i \delta(s)}\theta\!\left(s-4 M_\pi^2\right),
\end{align}
as 
\begin{align}\label{eq:discF}
    \operatorname{disc} \mathcal{F}(s) &=  2 i f_1(s) \sin \delta(s)\, e^{-i \delta(s)}\theta\!\left(s-4 M_\pi^2\right) \nonumber\\
    &\equiv  2 i(\mathcal{F}(s)+\hat{\mathcal{F}}(s)) \sin \delta(s)\, e^{-i \delta(s)}\theta\!\left(s-4 M_\pi^2\right),
\end{align}
where $\delta$ is the $\pi\pi$ $P$-wave phase shift and $\hat{\mathcal{F}}(s)$ is called the inhomogeneity of the integral equation, which contains the $s$-channel projection of the contributions of the $t$- and $u$-channels. 

Matching Eq.~\eqref{eq:KT_decomposition} with the $s$-channel partial-wave expansion Eq.~\eqref{eq:PW}, we obtain the relation between $\hat{\mathcal{F}}$ and angular averages over $\mathcal{F}$,
\begin{align}
    \hat{\mathcal{F}}(s)=3\left\langle\left(1-z_s^2\right) \mathcal{F}\right\rangle(s),\quad \left\langle z_s^n \mathcal{F}\right\rangle \equiv \frac{1}{2} \int_{-1}^1 \md z_s~z_s^n \mathcal{F}\left(\frac{3 s_0-s+z_s \kappa(s)}{2}\right). \label{eq:SVA}
\end{align}
On the other hand, Eq.~\eqref{eq:discF} has a solution~\cite{Babelon:1976kv}
\begin{align}\label{eq:mF_v1}
    \mF(s)=\Omega(s)\left\{a+\frac{s}{\pi} \int_{4 M_\pi^2}^{\infty} \frac{\md s^{\prime}}{s^{\prime}} \frac{\sin \delta\!\left(s^{\prime}\right) \hat{\mF}\left(s^{\prime}\right)}{\left|\Omega\left(s^{\prime}\right)\right|\left(s^{\prime}-s\right)}\right\},
\end{align}
where $\Omega(s)$ is the ($P$-wave) Omn\`es function calculated from $\delta(s)$ and $a$ is a (complex) subtraction or normalization constant. The number of subtractions is determined by the high-energy behavior of $\delta(s)$, which is assumed to asymptotically approach $\pi$ as $s\to +\infty$. 
Thus, a once-subtracted DR is needed for the convergence of Eq.~\eqref{eq:mF_v1}.\footnote{Since the subtraction constant $a$ actually serves as a normalization factor here, Refs.~\cite{JPAC:2020umo,JPAC:2023nhq,Garcia-Lorenzo:2025uzc} also refer to it as an unsubtracted dispersion relation for $\mF$.} 

Since $\hat{\mF}$ is linear in $\mF$ and the subtraction constant $a$ enters linearly in Eq.~\eqref{eq:mF_v1}, $\mF(s)$, $\hat{\mF}$, and $f_1$ can be decomposed as follows for numerical implementation:
\begin{align}\label{eq:KTeq}
\begin{aligned}
& \mathcal{F}(s)=a \mathcal{F}_a(s), \quad \hat{\mathcal{F}}(s)=a \hat{\mathcal{F}}_a(s),\quad f_1(s)=a\left(\mF_a(s)+\hat{\mathcal{F}}_a(s)\right)\equiv a\bar{\mF}(s), \\
& \mathcal{F}_a(s)=\Omega(s)\left\{1+\frac{s}{\pi} \int_{4 M_\pi^2}^{\infty} \frac{\md s^{\prime}}{s^{\prime}} \frac{\sin \delta\!\left(s^{\prime}\right) \hat{\mathcal{F}}_a\left(s^{\prime}\right)}{\left|\Omega\left(s^{\prime}\right)\right|\left(s^{\prime}-s\right)}\right\} ,\quad \hat{\mathcal{F}}_a(s)=3\left\langle\left(1-z_s^2\right) \mathcal{F}_a\right\rangle(s),
\end{aligned}
\end{align}
where $\mF_a(s)$ is called the KT basis function. 
Consequently, our result for the line shape of the TFF will be a pure prediction, up to the overall normalization constant $a$. 
Using data compiled by the Particle Data Group (PDG)~\cite{ParticleDataGroup:2024cfk}, its modulus $|a|$ can be fixed to reproduce the experimental $J/\psi\to 3\pi$ decay width [cf.\ Eq.~\eqref{eq:JpsiTo3pi_width}]. 
Since the overall phase $\operatorname{arg}(a)$ is not observable, we can simply set it to zero. 
In other words, only relative phases between different contributions to the TFF are physical: any nonzero $\operatorname{arg}(a)$ can be absorbed into a redefinition of the phases of the remaining weight factors introduced below.

The KT representation of the amplitude $f_1$ is given in Eq.~\eqref{eq:KTeq}. We denote $M_{\pm}^2=(M_\psi\pm M_\pi)^2$. The inhomogeneity $\hat{\mF}$ was derived in the physical region of the $s$-channel process $J/\psi\pi^0\to \pi^+\pi^-$, which corresponds to $s\geq M_+^2$ and $|z_s|\leq 1$. 
To obtain $\hat{\mF}$ in the physical region for the decay $J/\psi\to \pi^0\pi^+\pi^-$, $4M_\pi^2\leq s \leq M_-^2$, the right-hand side of $\hat{\mF}$ in Eq.~\eqref{eq:KTeq} must be analytically continued in the complex $s$ plane. 
Replacing the integration over $z_s$ with integration over $t$, the angular integral becomes
\begin{align}
    \left\langle z_s^n \mathcal{F}\right\rangle = \frac{1}{2} \int_{-1}^1 \mathrm{d} z_s~z_s^n \mathcal{F}\left(\frac{3 s_0-s+z_s \kappa(s)}{2}\right)=\frac{1}{\kappa(s)} \int_{t_{-}(s)}^{t_{+}(s)} \mathrm{d} t~\left(\frac{2 t-3 s_0+s}{\kappa(s)}\right)^n \mathcal{F}\left(t\right),
\end{align}
where
\begin{align}
    t_{ \pm}(s)=\frac{3 s_0-s \pm \kappa(s)}{2}.
\end{align}

\begin{figure}[t]
    \centering
    \includegraphics[width=.8\linewidth]{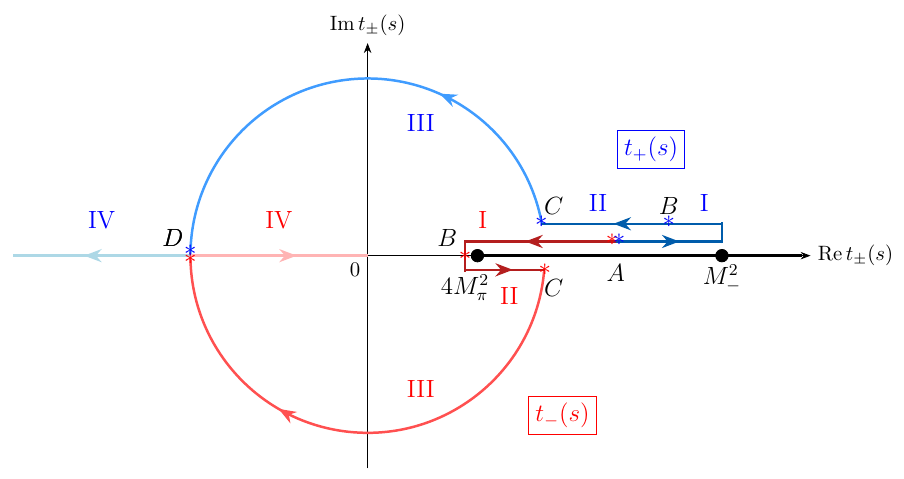}
    \caption{The KT path of $t_{ \pm}(s)$ [cf.\ Eq.~\eqref{eq:t+-}]. The trajectories start at the same point $A$ with $t_{\pm}(4 M_\pi^2)$ above the physical cut. At point $B$ with $t_{\pm}((M_\psi^2-M_\pi^2)/2)$, the $t_{-}$ contour reaches $4 M_\pi^2$ and moves below the cut. From point $C$ with $t_{ \pm}(M_-^2)$ onwards to point $D$ with $t_{ \pm}(M_+^2)$, $t_{-}$ and $t_{+}$ have the same real part but opposite imaginary parts. The trajectories finally meet at point $D$. From point $D$ onwards, both trajectories are real with $t_{-}$ moving towards zero and $t_{+}$ towards infinity.}
    \label{fig:KTpath}
\end{figure}

In the $s$-channel physical region, the integration limits $t_\pm$ lie on the negative real axis labeled region IV in Fig.~\ref{fig:KTpath} and do not overlap with the right-hand cut of $\mF$ for $s>4M_\pi^2$. 
As shown in Ref.~\cite{Kambor:1995yc}, analytic continuation to the decay region requires deforming the integration path in a way that does not cross the right-hand cut, as shown in Fig.~\ref{fig:KTpath}. 
The contour deformation of the KT equation has been extensively investigated in Refs.~\cite{Bronzan:1963mby, Kambor:1995yc}.\footnote{Note that there is another powerful, albeit more technical method for the analytic continuation of the KT equation~\cite{Aitchison:1966lpz,Gasser:2018qtg}. In that method, contour deformation is applied not to the angular integral but to the dispersion integral of $\mF$ shown in Eq.~\eqref{eq:KTeq}. This approach avoids the spurious singularity at the pseudothreshold $(M_\psi-M_\pi)^2$ that arises in the traditional solution strategy of the KT equation, provided that the analytic continuation of the phase shifts is handled carefully. Alternatively, $\sin\delta(s')/|\Omega(s')|$ in Eq.~\eqref{eq:SVA} can be rewritten in terms of the partial wave directly~\cite{Niehus:2021iin}.} Following the analysis of the analytic continuation of the triangle topology diagram in perturbation theory~\cite{Bronzan:1963mby}, the proper analytic continuation of $t_\pm$ is obtained by employing the $+i\epsilon$ prescription to the initial-state energy, i.e., $t_\pm\big(s,M_\psi^2\big)\to t_\pm\big(s,M_\psi^2+i\epsilon\big)$, which yields
\begin{align}
    t_{ \pm}\left(s, M_\psi^2+i \epsilon\right)=t_{ \pm}\left(s, M_\psi^2\right)+i \epsilon \frac{\partial t_{ \pm}\big(s, M_\psi^2\big)}{\partial M_\psi^2}.
\end{align}
We thus obtain the explicit expressions~\cite{Kambor:1995yc, Niecknig:2012sj}\footnote{We use the analytic continuation of $\kappa$~\cite{Bronzan:1963mby}: $\kappa(s)=\sigma_\pi(s)\sqrt{\left(M_+^2-s\right)}\sqrt{\left(M_-^2-s\right)}$.}
\begin{align}\label{eq:t+-}
\begin{aligned}
& 2 t_{+}(s)= \begin{cases}3 s_0-s+|\kappa(s)|+i \epsilon, & s \in\left[4 M_\pi^2, M_-^2\right], \quad \rm{I\& II} \\
3 s_0-s+i|\kappa(s)|, & s \in[M_-^2, M_+^2], \quad \rm{III} \\
3 s_0-s-|\kappa(s)|, & s \in[M_+^2, \infty), \quad \rm{IV} \end{cases} \\
& 2 t_{-}(s)= \begin{cases}3 s_0-s-|\kappa(s)|+i \epsilon, & s \in\left[4 M_\pi^2, \frac{M_\psi^2-M_\pi^2}{2}\right], \quad \rm{I} \\
3 s_0-s-|\kappa(s)|-i \epsilon, & s \in\left[\frac{M_\psi^2-M_\pi^2}{2}, M_-^2\right], \quad \rm{II} \\
3 s_0-s-i|\kappa(s)|, & s \in[M_-^2, M_+^2], \quad \rm{III} \\
3 s_0-s+|\kappa(s)|, & s \in[M_+^2, \infty), \quad \rm{IV} \end{cases}
\end{aligned}
\end{align}
where \rm{I}, \rm{II}, \rm{III}, and \rm{IV} are the regions labeled in Fig.~\ref{fig:KTpath}.

\begin{figure}[t]
  \centering
  \includegraphics[width=1\textwidth]{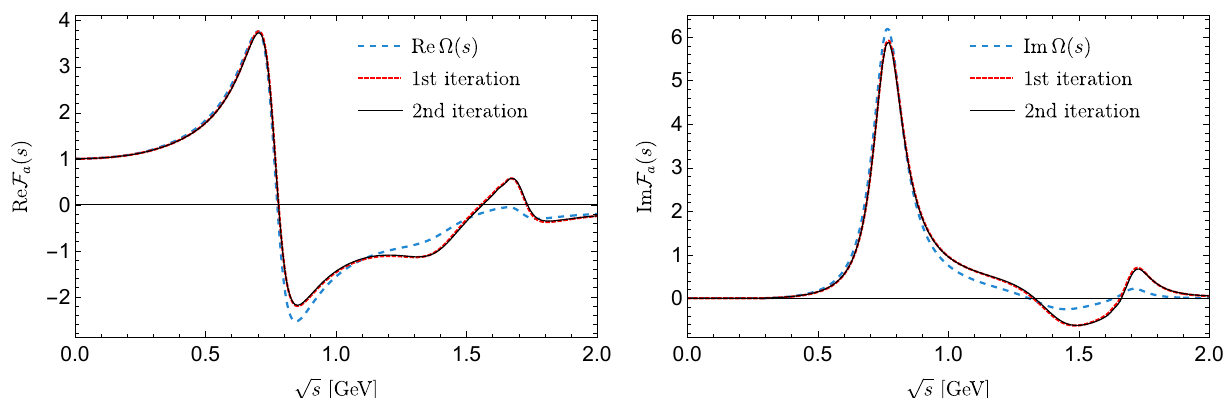}
  \caption{Convergence behavior of the iterative procedure for the real (left panel) and imaginary (right panel) parts of the $P$-wave basis solution $\mathcal{F}_a(s)$ for $J/\psi\to 3\pi$ using solution~1 of the phase shift as input.}
  \label{fig:F}
\end{figure}

We solve the integral equation in Eq.~\eqref{eq:KTeq} by numerical iterative procedure.\footnote{To obtain numerical solutions of the KT equation, we refer to Refs.~\cite{Schneider:2012nng,Niecknig:2015ija,Niecknig:2016fva,Albaladejo:2019huw,Stamen:2022eda}. We are not aware of a general proof of convergence for this iterative procedure; however, in our numerical implementation for the present kernel and input phases, the solution stabilizes already after the second iteration.} 
The final results of $\mathcal{F}_a$ are shown in Fig.~\ref{fig:F}. The difference between the Omn\`es function and $\mF_a(s)$ serves as a measure of the importance of crossed-channel rescattering effects. We clearly see that, compared to the $\omega,\phi\to 3\pi$ results in Refs.~\cite{Niecknig:2012sj, Danilkin:2014cra,JPAC:2020umo}, the crossed-channel rescattering effects are quite small due to the mass hierarchy $M_\psi\gg M_{\omega,\phi}$. 
Nevertheless, the effects need to be included to obtain a satisfactory description of the fine structure near the $\rho$ and $\rho'$ resonances~\cite{JPAC:2023nhq}.

\begin{table}[t]
    \centering
    \caption{Input parameters used throughout this work. Regarding the treatment of resonances, the quoted masses and widths are reaction dependent, referring to Breit-Wigner (BW) parameters rather than model-independent pole parameters. We use the $\rho(770)$ parameters from the PDG averaged values~\cite{ParticleDataGroup:2024cfk} for the charged channel.}
\begin{tabular}{llr}
\hline\hline
Quantity & Value & Ref. \\
\hline $M_\psi$ & $3096.900(6)~\mathrm{MeV}$ & \cite{ParticleDataGroup:2024cfk} \\
$\Gamma_\psi$ & $92.6(1.7)~\mathrm{keV}$ & \cite{ParticleDataGroup:2024cfk} \\
$\mathcal{BR}(J / \psi \to\pi^{+} \pi^{-} \pi^0)$ & $2.00(7)\%$ & \cite{ParticleDataGroup:2024cfk} \\
$\mathcal{BR}(J / \psi \to \pi^0 \gamma)$ & $3.39(8)\times 10^{-5}$ & \cite{ParticleDataGroup:2024cfk} \\
$\epsilon_{\rho\omega}$ & $1.99(2)$ & \cite{Leplumey:2025kvv} \\
$M_\omega$ & $782.66(13)~\mathrm{MeV}$ & \cite{ParticleDataGroup:2024cfk} \\
$\Gamma_\omega$ & $8.68(13)~\mathrm{MeV}$ & \cite{ParticleDataGroup:2024cfk} \\
$\mathcal{BR}\left(\omega \rightarrow e^{+} e^{-}\right)$ & $7.41(19)\times 10^{-5}$ & \cite{ParticleDataGroup:2024cfk} \\
$\mathcal{BR}(J/\psi\to\omega\pi^0)$ & $4.5(5)\times 10^{-4}$ & \cite{ParticleDataGroup:2024cfk} \\
$M_\phi$ & $1019.460(16)~\mathrm{MeV}$ & \cite{ParticleDataGroup:2024cfk} \\
$\Gamma_\phi$ & $4.249(13)~\mathrm{MeV}$ & \cite{ParticleDataGroup:2024cfk} \\
$\mathcal{BR}\left(\phi \rightarrow e^{+} e^{-}\right)$ & $2.964(33)\times 10^{-4}$ & \cite{ParticleDataGroup:2024cfk} \\
$\mathcal{BR}(J/\psi\to\phi\pi^0)$ & $3\times 10^{-6}$ or $1\times 10^{-7}$ & \cite{ParticleDataGroup:2024cfk} \\
$M_\rho$ & $775.11(34)~\mathrm{MeV}$ & \cite{ParticleDataGroup:2024cfk} \\
$M_{\rho^\prime}$ & $1465(25)~\mathrm{MeV}$ & \cite{ParticleDataGroup:2024cfk} \\
$\Gamma_{\rho^\prime}$ & $400(60)~\mathrm{MeV}$ & \cite{ParticleDataGroup:2024cfk} \\
$\mathcal{B R}\left(J / \psi \rightarrow \rho^{\prime} \pi\right) \mathcal{B R}\left(\rho^{\prime} \rightarrow 2 \pi\right)$ & $2.2(1.1)\times 10^{-4}$ & \cite{ParticleDataGroup:2024cfk} \\
$\mathcal{B R}\left(\rho^{\prime} \rightarrow 2 \pi\right)$ & $6\%$ & \cite{Zanke:2021wiq} \\
\hline\hline
\end{tabular}
    \label{tab:input_params}
\end{table}

In particular, we can calculate the two-pion contribution~\eqref{eq:2piDisc} to the real-photon transition $J/\psi\to \pi^0\gamma$ using the sum rule~\eqref{eq:sum rule_v1}. The single unknown parameter $a$ only affects the overall normalization of the amplitude and can be fixed from the $J/\psi \to \pi^{+} \pi^{-} \pi^0$ decay width, obtained integrating the distribution Eq.~\eqref{eq:JpsiTo3pi_width}. 
Using the PDG averaged value in Table~\ref{tab:input_params}, one finds $|a|=0.043(2)~\mathrm{GeV}^{-3}$, where the main source of uncertainty comes from the discrepancy between the two input solutions for the phase shift in Fig.~\ref{fig:phase}. Using the experimental branching fraction for $J/\psi \rightarrow \pi^0 \gamma$~\cite{ParticleDataGroup:2024cfk}, together with Eq.~\eqref{eq:pi0gamma_decay}, yields 
\begin{align}
    \left|f_{\psi \pi^0}(0)\right|=6.0(3) \times 10^{-4}~\mathrm{GeV}^{-1}, \label{eq:normalization_psipi0}
\end{align}
whereas the sum rule gives
\begin{align}\label{eq:normalization_2pi}
    \left|f_{\psi \pi^0}^{(2\pi)}(0)\right|=4.8(2) \times 10^{-4}~\mathrm{GeV}^{-1},
\end{align}
which is consistent with previous KT equation analyses~\cite{Kubis:2014gka,JPAC:2023nhq}. Note that the phase $\left|\arg \left(f_{\psi \pi^0}^{(2\pi)}(0)\right)\right|$ is less than $3^\circ$. Therefore, for simplicity, we treat $f_{\psi \pi^0}^{(2 \pi)}(0)$ as a real number in the following discussion.

\subsection{Effective three-pion intermediate states: $\rho$--$\omega$ mixing effects}

Currently lattice QCD has arrived at the point of being able to extract $\rho$ and $\omega$ simultaneously~\cite{Yan:2024gwp}. While at the moment there are no isospin-breaking effects included, the next generation computations will address these questions from QCD. $\rho$--$\omega$ mixing occurs because both the $\rho^0$ and $\omega$ mesons are not isospin eigenstates. 
Consequently, $\omega$ has a small, but nonvanishing branching fraction to $2\pi$, which leads to $\rho$--$\omega$ interference pattern in the $J/\psi\to \pi^0 \gamma^*$ TFF as clearly observed in the data measured by BESIII~\cite{BESIII:2025xjh}. 
This effect is enhanced by a factor $M_\omega/\Gamma_\omega\approx 90$ through the presence of the narrow $\omega$-resonance propagator and therefore must be included to obtain a realistic line shape.\footnote{The absence of significant enhancement on the left shoulder of the $\omega$ resonance has been established in studies of nucleon electromagnetic form factors~\cite{Bernard:1996cc,Kaiser:2019irl}. This is due to the different phase space factors, which are $(s-4M_\pi^2)^{3/2}$ and $(s-9M_\pi^2)^4$ for the isovector and isoscalar cases, respectively. This justifies the standard dispersive approach of taking only the $\omega$ pole as the lowest-lying singularity in the isoscalar spectral function.} 
This situation implies that the $2\pi$ and $3\pi$ channels are coupled and cannot be investigated separately.\footnote{Interestingly, the inverse mixing effect of the $\rho$ affecting the $3\pi$ spectrum has now also been established~\cite{BABAR:2021cde,Hoferichter:2023bjm}, although this obviously lacks the enhancement of a narrow resonance and rather results in a broad background.} 

The $\rho$--$\omega$ mixing in the pion vector form factor $F_\pi^V$ requires careful treatment. If one naively introduces an isospin-breaking factor as
\begin{align}
    F_\pi^V(s)\to \left(1+\epsilon_{\rho \omega} \frac{s}{M_\omega^2-s-i M_\omega \Gamma_\omega}\right)F_\pi^V(s) 
\end{align}
in $F_\pi^V$ within the discontinuity relation~\eqref{eq:2piDisc}, the phase on the right-hand side of the above equation would not cancel that of $f_1^*(s)$ due to the $\omega$ propagator, in contrast to the case without $\rho$--$\omega$ mixing in Eq.~\eqref{eq:2piDisc}.\footnote{This statement is to be understood in the physical scattering region: after analytic continuation to the decay region, crossed-channel effects in $f_1$ already lead to a complex discontinuity, which, however, is a physical effect and not due to an invalid approximation.}
Here, $\epsilon_{\rho \omega}$ is a $\rho$--$\omega$ mixing parameter that can be extracted from $e^+e^-\to\pi^+\pi^-$ data~\cite{Leplumey:2025kvv}. 
Such behavior would occur if only the $2 \pi$ discontinuity mixed with $\omega$ was considered. However, there should also be an isospin-violating $\rho$ contribution to the $3 \pi$ discontinuity, which will mix with the $2 \pi$ channel. Only when both contributions are accounted for can a consistent spectral function be obtained~\cite{Hoferichter:2016duk}.

\begin{figure}[t]
    \centering
    \includegraphics[width=.6\linewidth]{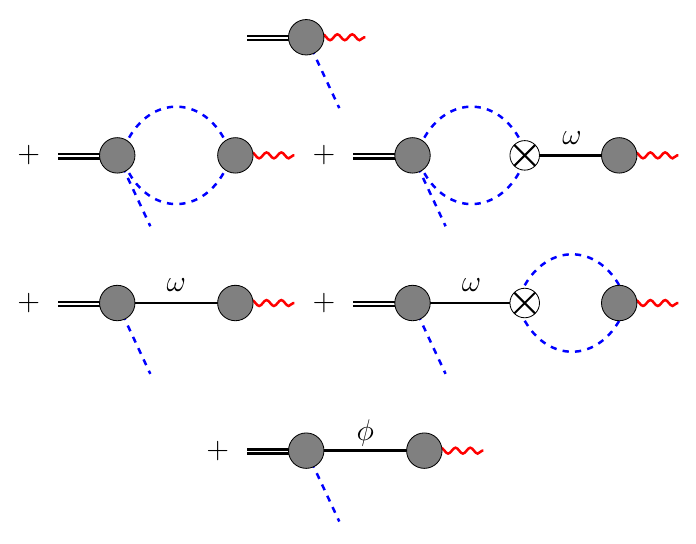}
    \caption{Pictorial representation of Eq.~\eqref{eq:2pi3pi_DR}.
    We show the diagrams corresponding to the different terms in the right-hand side of the equation in order.
    Dashed blue lines denote pions, full black lines the $\omega$ or $\phi$ propagators, filled vertices show the form factors, those marked with a cross denote the isospin-violating $\rho$--$\omega$ mixings.}\label{fig:rho-omega_mixing}
\end{figure}

Here, we systematically employ a dispersive formalism that enables a consistent implementation of $\rho$--$\omega$ mixing in both isoscalar and isovector contributions. This approach has been derived and applied to the TFF of $\eta^{\prime}$ to $\gamma^\ast \gamma$ in Ref.~\cite{Holz:2022hwz} using a coupled-channel two-potential formalism~\cite{Nakano:1982bc} (see also Refs.~\cite{Hanhart:2012wi,Ropertz:2018stk,VonDetten:2021rax,Heuser:2024biq} for applications to light-meson form factors and resonances). Applying this approach to the problem at hand, the dispersive representation of the two-pion and three-pion contributions to $f_{\psi\pi^0}^{(2\pi,3\pi)}(s)$ is given by
\begin{align}\label{eq:2pi3pi_DR}
    f^{(2\pi,3\pi)}_{\psi\pi^0}(s)= &\, f^{(2\pi)}_{\psi\pi^0}(0) \nonumber\\
& +\frac{s}{96 \pi^2} \int_{4 M_\pi^2}^{\infty} \mathrm{d} s^{\prime} \frac{\sigma_\pi^3\left(s^{\prime}\right) F_\pi^{V *}\left(s^{\prime}\right)f_1\left(s^{\prime}\right)}{s^{\prime}-s-i \epsilon} \left[1+\frac{\tilde{\epsilon}_{\rho \omega} s}{M_\omega^2-s-i M_\omega \Gamma_\omega}\right] \nonumber\\
& +\frac{w_{\psi \omega \pi} s}{M_\omega^2-s-i M_\omega \Gamma_\omega}\left[1+\frac{\tilde{\epsilon}_{\rho \omega} s}{48 \pi^2 g_{\omega \gamma}^2}\int_{4 M_\pi^2}^{\infty} \mathrm{d} s^{\prime} \frac{\sigma_\pi^3\left(s^{\prime}\right) |F_\pi^{V}\left(s^{\prime}\right)|^2}{s^{\prime}\!\left(s^{\prime}-s-i \epsilon\right)}\right] \nonumber\\
& +\frac{w_{\psi \phi \pi} s}{M_\phi^2-s-i M_\phi \Gamma_\phi},
\end{align}
where the $\omega$--photon coupling $g_{\omega\gamma}$ can be extracted from the corresponding electron--positron decay width for a vector meson $V$:\footnote{Here, we use the VMD effective Lagrangian $\mathcal{L}_{V \gamma}=e M_V^2 A^\mu g_{V \gamma} V_\mu$ and the phase convention is chosen to be $\operatorname{sgn}g_{\omega\gamma}=+1$.}
\begin{align}
    \Gamma\left(V \rightarrow e^{+} e^{-}\right)=\frac{4 \pi \alpha^2 g_{V \gamma}^2 M_V}{3}.
\end{align}
The value of the coupling is
\begin{align}
    g_{\omega \gamma}=\sqrt{\frac{3 \Gamma\left(\omega \rightarrow e^{+} e^{-}\right)}{4 \pi \alpha^2 M_\omega}}=0.0606(9).
\end{align}
A schematic illustration of Eq.~\eqref{eq:2pi3pi_DR} is shown in Fig.~\ref{fig:rho-omega_mixing}.

The mixing parameter $\tilde\epsilon_{\rho\omega}$ can be decomposed into QCD $\mathcal{O}\left(m_u-m_d\right)$ and quantum electrodynamics (QED) $\mathcal{O}\left(e^2\right)$ contributions,
as discussed in Refs.~\cite{Gasser:1982ap,Urech:1995ry}; see also Refs.~\cite{Holz:2022hwz,Colangelo:2022prz,Dias:2024zfh}. One has 
\begin{align}
    \tilde{\epsilon}_{\rho\omega}=\frac{2}{3 R} \frac{M_{K^*}-M_\rho}{M_\rho}-e^2 g_{\omega \gamma}^2, \quad R=\frac{m_s-\hat{m}}{m_d-m_u}, \quad \hat{m}=\frac{m_u+m_d}{2},
\end{align}
where the electromagnetic component is expressed in terms of the $\omega$-photon coupling, since the photon has both isovector and isoscalar components and can therefore mediate the mixing between $\rho$ and $\omega$. 
However, it yields a one-particle-reducible correction that is divided out when the QED vacuum polarization is removed from the $e^{+} e^{-} \rightarrow \pi^{+} \pi^{-}$ cross sections, which is usually done in experimental analyses; the $\rho$--$\omega$ mixing parameter extracted in this case is denoted as $\epsilon_{\rho\omega}$. Consequently, the parameter $\tilde{\epsilon}_{\rho\omega}$ is related to $\epsilon_{\rho\omega}$ from the $e^{+} e^{-} \rightarrow \pi^{+} \pi^{-}$ cross sections by
\begin{align}
    \tilde{\epsilon}_{\rho\omega}=\epsilon_{\rho\omega}-e^2g_{\omega\gamma}^2=1.99(2)\times10^{-3}-0.34(1)\times10^{-3}=1.65(2)\times10^{-3}.
\end{align}

The isoscalar contributions of $\omega$ and $\phi$ to the TFF are proportional to the weight factors $w_{\psi\omega\pi}$ and $w_{\psi\phi\pi}$, respectively, which can have nonzero small phases. One source of these phases arises from the mixing of virtual-photon intermediate states with isoscalar vector mesons, induced by QED effects. To estimate the magnitudes of the weight factors $w_{\psi\omega\pi}$ and $w_{\psi\phi\pi}$, we match the expressions for the differential width of the decay $J/\psi \to \pi^0 e^+ e^-$ at the narrow isoscalar vector meson poles,
\begin{align}\label{eq:BW}
    \left.\frac{\mathrm{d} \Gamma_{\psi \rightarrow \pi^0 e^{+} e^{-}}}{\mathrm{~d} s}\right|_{s\to M_V^2}=\frac{\Gamma_\psi \mathcal{BR}(J/\psi\to V\pi^0)\mathcal{BR}(V\to e^+ e^-)}{\pi \Gamma_V M_V},
\end{align}
where $m_e$ is always set to 0 unless otherwise stated and $V=\omega$ or $\phi$. Here, we adopt the effective Lagrangian $\mathcal{L}_{V_1V_2\pi^0}=g_{V_1V_2\pi}\epsilon_{\mu\nu\alpha\beta}\partial^\mu V_1^\nu \partial^\alpha V_2^\beta \pi^0$ that results in the $V_1\to V_2\pi^0$ decay width
\begin{align}\label{eq:V->Vpi}
    \Gamma_{V_1\to V_2\pi}=\frac{g^2_{V_1V_2\pi}}{96\pi M_{V_1}^3}\lambda^{\frac{3}{2}}\!\left(M_{V1}^2,M_{V_2}^2,M_\pi^2\right).
\end{align}
We also obtain from Eq.~\eqref{eq:diff_Gamma},
\begin{align}
    \left.\frac{\mathrm{d} \Gamma_{\psi \rightarrow \pi^0 e^{+} e^{-}}}{\mathrm{~d} s}\right|_{s\to M_V^2}=\frac{\alpha^2}{72\pi M_V^3 \Gamma_V^2}\lambda^{\frac{3}{2}}\!\left(M_V^2,M_\psi^2,M_\pi^2\right)\left|w_{\psi V\pi}\right|^2,
\end{align}
where we assume isoscalar vector resonance pole dominance,
\begin{align*}
    \left.f_{\psi \pi}(s)\right|_{s\to M_V^2}=\frac{w_{\psi V \pi} s}{M_V^2-s-i M_V \Gamma_V}.
\end{align*}
The branching ratios or partial decay widths tabulated by the PDG~\cite{ParticleDataGroup:2024cfk} are then sufficient to determine the magnitude of the weight factors. The final result is 
\begin{align}\label{eq:modulus_w}
    |w_{\psi V\pi}| & =\sqrt{\frac{72 M_\psi^3 \Gamma_\psi \Gamma_V \mathcal{BR}(V\to e^+e^-)\mathcal{BR}(J/\psi\to V\pi^0)}{\alpha^2 M_V \lambda^\frac{3}{2}\left(M_\psi^2,M_V^2,M_\pi^2\right)}} \nonumber\\
    & =
    \begin{cases}
    4.38(24)\times10^{-5}~\text{GeV}^{-1},\quad V=\omega , \\
    4.71(3)\times10^{-6}~\text{GeV}^{-1}\ \text{or}\ 8.61(5)\times10^{-7}~\text{GeV}^{-1},\quad V=\phi .
    \end{cases}
\end{align}
This method is only valid for narrow intermediate resonances and appears to be quite good for $\omega$ and $\phi$. In addition, this approach neglects possible non-negligible phases, especially in the case of $\omega$.

Equation~\eqref{eq:2pi3pi_DR} is a once-subtracted dispersion relation constructed from a spectral function whose double discontinuity vanishes. As a consistency check, we show that the imaginary parts in the weight factors in Eq.~\eqref{eq:2pi3pi_DR} are indeed compatible with unitarity. Applying the Cutkosky cutting rules (or the discontinuity calculus~\cite{Jing:2025qmi}) to Eq.~\eqref{eq:2pi3pi_DR}, we have
\begin{align}
\left.\frac{1}{2 i} \operatorname{disc} f_{\psi\pi^0}(s)\right|_{2 \pi}= & \left[1+\frac{\tilde{\epsilon}_{\rho\omega}s}{M_\omega^2-s-i \epsilon}\right]\operatorname{Im}\Sigma(s)+\frac{f_{\psi\pi^0}(0)w_{\psi\omega\pi}s}{M_\omega^2-s-i \epsilon} \frac{\tilde{\epsilon}_{\rho\omega}}{g_{\omega\gamma}^2}\frac{\operatorname{Im}\Sigma_\pi(s)}{s}, \nonumber\\
\left.\frac{1}{2 i} \operatorname{disc} f_{\psi\pi^0}(s)\right|_{3 \pi}= &\, \pi s \delta\!\left(s-M_\omega^2\right)\tilde{\epsilon}_{\rho\omega}\Sigma(s)+f_{\psi\pi^0}(0) w_{\psi\omega\pi}^* \pi s \delta\!\left(s-M_\omega^2\right)\left[1+\frac{\tilde{\epsilon}_{\rho\omega}}{g_{\omega\gamma}^2}\frac{\Sigma_\pi(s)}{s}\right]^* \nonumber\\
& +f_{\psi\pi^0}(0) w_{\psi\phi\pi}^* \pi s \delta\!\left(s-M_\phi^2\right), \nonumber\\
\left.\frac{1}{2 i} \operatorname{disc} f_{\psi\pi^0}(s)\right|_{\gamma}= &\, f_{\psi\pi^0}(0)\operatorname{Im} w_{\psi\omega\pi} \frac{s}{M_\omega^2-s-i \epsilon}\left[1+\frac{\tilde{\epsilon}_{\rho\omega}}{g_{\omega\gamma}^2}\frac{\Sigma_\pi(s)}{s}\right]^* \nonumber\\
& +f_{\psi\pi^0}(0)\operatorname{Im} w_{\psi\phi\pi} \frac{s}{M_\phi^2-s-i \epsilon} 
\end{align}
in the narrow-width limit $\Gamma_{\omega/\phi}\to 0$, where
\begin{align}
    \Sigma(s) & =\frac{s}{96 \pi^2} \int_{4 M_\pi^2}^{\infty} \mathrm{d} s^{\prime} \frac{\sigma_\pi^3\left(s^{\prime}\right) F_\pi^{V *}\left(s^{\prime}\right)f_1\left(s^{\prime}\right)}{s^{\prime}-s-i \epsilon}, \\
    \Sigma_\pi(s) & =\frac{s^2}{48 \pi^2} \int_{4 M_\pi^2}^{\infty} \mathrm{d} s^{\prime} \frac{\sigma_\pi^3\left(s^{\prime}\right) |F_\pi^{V}\left(s^{\prime}\right)|^2}{s^\prime(s^{\prime}-s-i \epsilon)}.
\end{align}
These equations are consistent only if the sum of these discontinuities is purely imaginary. Collecting all terms, this nontrivial consistency check is indeed satisfied,
\begin{align}
    \operatorname{Im}\left[\left.\frac{1}{2 i} \operatorname{disc} f_{\psi\pi^0}(s)\right|_{2 \pi}+\left.\frac{1}{2 i} \operatorname{disc} f_{\psi\pi^0}(s)\right|_{3 \pi}+\left.\frac{1}{2 i} \operatorname{disc} f_{\psi\pi^0}(s)\right|_{\gamma}\right]=0,
\end{align}
as long as $f_1(s)$ is within the scattering region.

\subsection{Effective four-pion intermediate states: $\rho^\prime(1450)$}
\label{sec:rho'}

The branching fractions of $J/\psi$ into several final states with more than three pions are actually as large as that into $\pi^+\pi^-\pi^0$. This suggests that inelastic contributions to the $J / \psi \rightarrow \pi^0 \gamma^\ast$ TFF, arising from discontinuities due to four or more pions, could play a sizable role. 
In the simplest model approach, we incorporate a $\rho^{\prime}(1450)$ resonance into the $J / \psi \rightarrow \pi^0 \gamma^\ast$ TFF as an approximation for the potential effects of multipion intermediate states. 
This method is based on the approach outlined in Ref.~\cite{Kubis:2014gka}. However, since Ref.~\cite{Kubis:2014gka} contains some ambiguities and typographical errors in the $\rho^\prime(1450)$ contribution,\footnote{The product of couplings $g_{J / \psi \rho^{\prime} \pi} g_{\rho^{\prime} \gamma}$ is missing in  the expression of $\operatorname{disc} f_{\psi \pi^0}^{\rho^{\prime}}(s)$ in Eq.~(14) of Ref.~\cite{Kubis:2014gka}; see Eq.~\eqref{eq:rhoprime_DR}. In addition, the exponent $7/2$ in the factor $\Big[\frac{s-16 M_\pi^2}{M_{\rho^{\prime}}^2-16 M_\pi^2}\Big]^{7 / 2}$ in the same equation of Ref.~\cite{Kubis:2014gka} should be corrected to $9/2$; see Eq.~\eqref{eq:width_v1}.} we present a more comprehensive analysis here.
This effective $\rho^\prime$ should not be confused with the elastic imprint of higher resonances already encoded in solution~1 for the phase of $F_\pi^V$. Here it is introduced solely as an estimate of genuinely inelastic multipion strength, so no double counting occurs.

Note that even for channels for which no rigorous spectral functions as in Eq.~\eqref{eq:2piDisc} are known, an ansatz for the full momentum dependence with good analytic properties can be constructed by replacing the energy-dependent BW propagator by a dispersion relation~\cite{Aitchison:1964rwb}:
\begin{align}\label{eq:rhoprime_DR}
    f_{\psi \pi^0}^{(\rho^\prime)}(s)=\frac{1}{2 \pi i} \int_{4 M_\pi^2}^{\infty} \mathrm{d} s^{\prime}~\frac{\operatorname{disc} f_{\psi \pi^0}^{(\rho^\prime)}\left(s^{\prime}\right)}{s^{\prime}-s-i \epsilon},\quad \frac{\operatorname{disc}}{2i} f_{\psi \pi^0}^{(\rho^{\prime})}(s)= & \frac{M_{\rho^\prime}^2 g_{J/\psi\rho^\prime \pi} g_{\rho^\prime\gamma} \sqrt{s} \Gamma_{\rho^{\prime}}(s)}{\left(M_{\rho^{\prime}}^2-s\right)^2+s \Gamma_{\rho^{\prime}}^2(s)},
\end{align}
where $|g_{\rho^\prime\gamma}|=0.0752$ is taken from the VMD model estimation~\cite{Zanke:2021wiq}.\footnote{Note that the coupling $g_{\rho^\prime\gamma}$ we use here corresponds to $1/g_{\rho^\prime\gamma}$ in Ref.~\cite{Zanke:2021wiq}.} The modulus of the effective coupling constant $g_{J/\psi \rho^\prime \pi}$ should be fixed from the decay width $J/\psi\to \rho^\prime \pi^0$ via Eq.~\eqref{eq:V->Vpi}. 
Unfortunately, the branching fraction of $J / \psi \rightarrow \rho^{\prime} \pi^0$ is not available. We estimate it using a cascade decay relation $\mathcal{BR}(J/\psi\to\rho^\prime\pi\to \pi^+\pi^-\pi^0)=\mathcal{BR}(J/\psi\to \rho^\prime \pi)\mathcal{BR}(\rho^\prime \to 2\pi)$. The branching ratio $\mathcal{BR}(\rho^{\prime} \to 2 \pi)$ can be obtained from the VMD model estimation, $\mathcal{BR}(\rho^{\prime} \to 2 \pi)=6\%$ ($\mathcal{BR}(\rho^{\prime} \to \omega\pi)=94\%$)~\cite{Zanke:2021wiq}, which was obtained by assuming that the main decay modes of $\rho^{\prime}$ are only $2 \pi$ and $\omega \pi$, while neglecting other significant contributions from, e.g., $\rho^{\prime} \rightarrow a_1 \pi$. Based on this, we obtain the branching ratio $\mathcal{BR}(J/\psi\to \rho^\prime \pi)=3.7(1.8)\times10^{-3}$ with the effective coupling constant $\left|g_{J/\psi\rho^\prime \pi}\right|=2.73^{+0.61}_{-0.80}\times 10^{-3}~\rm{GeV}^{-1}$.

For the energy-dependent width $\Gamma_{\rho^\prime}(s)$, we consider two different parameterizations. First, we assume that the decay channel $\rho^\prime \rightarrow 4\pi$ is dominant and thus adopt the near-threshold behavior of the four-pion phase space~\cite{Leutwyler:2002hm}. 
Second, we construct $\Gamma_{\rho^\prime}(s)$ from the decay channels $\rho^\prime \rightarrow \omega\pi(\omega \rightarrow 3\pi)$ and $\rho^\prime \rightarrow \pi\pi$, neglecting other contributions from, e.g., $\rho^\prime \rightarrow a_1\pi(a_1 \rightarrow 3\pi)$. 
This serves both as a consistency check for the assumption used in the above estimate of the effective coupling $|g_{J/\psi\rho^\prime \pi}|$ and as a theoretical uncertainty estimate for the sizable width effects of $\rho^\prime$. These parameterizations read~\cite{Zanke:2021wiq}
\begin{align}\label{eq:width_v1}
    \Gamma_{\rho^{\prime}}^{(4 \pi)}\left(s\right)=\theta\!\left(s-16 M_\pi^2\right) \frac{\gamma_{\rho^{\prime} \rightarrow 4 \pi}\left(s\right)}{\gamma_{\rho^{\prime} \rightarrow 4 \pi}\left(M_{\rho^{\prime}}^2\right)} \Gamma_{\rho^{\prime}}, \quad \gamma_{\rho^{\prime} \rightarrow 4 \pi}\left(s\right)=\frac{\left(s-16 M_\pi^2\right)^\frac{9}{2}}{s^2},
\end{align}
where $\Gamma_{\rho^{\prime}}$ is the total decay width of $\rho^{\prime}$, and
\begin{align}\label{eq:width_v2}
    \Gamma_{\rho^{\prime}}^{(\omega \pi, \pi \pi)}\left(s\right) =&\, \theta\!\left(s-\left(M_\omega+M_\pi\right)^2\right) \frac{\gamma_{\rho^{\prime} \rightarrow \omega \pi}\left(s\right)}{\gamma_{\rho^{\prime} \rightarrow \omega \pi}\left(M_{\rho^{\prime}}^2\right)} \Gamma_{\rho^{\prime} \rightarrow \omega \pi} \nonumber\\
    & +\theta\!\left(s-4 M_\pi^2\right) \frac{\gamma_{\rho^{\prime} \rightarrow \pi \pi}\left(s\right)}{\gamma_{\rho^{\prime} \rightarrow \pi \pi}\left(M_{\rho^{\prime}}^2\right)} \frac{M_{\rho'}^2-4 M_\pi^2+4 p_R^2}{s-4 M_\pi^2+4 p_R^2}\frac{\sqrt{s}}{M_{\rho'}} \Gamma_{\rho^{\prime} \rightarrow \pi \pi},
\end{align}
with $p_R=202.4$~MeV in the Blatt--Weisskopf barrier factor, and
\begin{align}
    \gamma_{\rho^{\prime} \rightarrow \omega \pi}\left(s\right)=\frac{\lambda^\frac{3}{2}\left(s, M_\omega^2, M_\pi^2\right)}{s^\frac{3}{2}}, \quad \gamma_{\rho^{\prime} \rightarrow \pi \pi}\left(s\right)=\frac{\left(s-4 M_\pi^2\right)^\frac{3}{2}}{s}.
\end{align}

\begin{figure}[tb]
    \centering
    \includegraphics[width=0.5\linewidth]{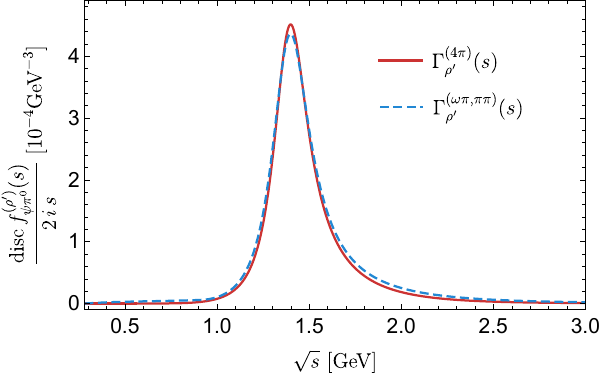}
    \caption{The discontinuities $\operatorname{disc}f^{(\rho^\prime)}_{\psi\pi^0}/(2\,i\,s)$ corresponding to two types of energy-dependent widths, cf.\ Eq.~\eqref{eq:width_v1} and Eq.~\eqref{eq:width_v2}.}
    \label{fig:discRhoPrime}
\end{figure}
We present the line shapes of the discontinuities $\operatorname{disc}f^{(\rho^\prime)}_{\psi\pi^0}/(2\,i\,s)$ corresponding to the two types of energy-dependent widths in Fig.~\ref{fig:discRhoPrime}. Both results show excellent agreement near the $\rho^\prime$ peak, further validating the consistency of our estimate. The curve for $\Gamma_{\rho^\prime}^{(\omega\pi, \pi\pi)}$ is slightly higher than that for $\Gamma_{\rho^\prime}^{(4\pi)}$ because the former includes a small (6\%) $2\pi$ contribution in its width. 
We use $\Gamma_{\rho^\prime}^{(4\pi)}$ as the central input in the following analysis.

Using the sum rule of Eq.~\eqref{eq:sum rule_v1}, we obtain
\begin{align}
    \left|f^{(\rho^\prime)}_{\psi \pi^0}(0)\right|=1.3^{+0.3}_{-0.4}\times 10^{-4}~\mathrm{GeV}^{-1}.
\end{align}
Comparing with Eq.~\eqref{eq:normalization_psipi0}, we conclude that the $\rho^\prime(1450)$ intermediate state alone contributes about 20\% of the sum rule for the TFF normalization. Recalling Eq.~\eqref{eq:normalization_2pi}, this seems to suggest that the contributions from both the $2\pi$ intermediate states and the $\rho^{\prime}$ fully saturate the sum rule for the TFF normalization (note that the effect of $\rho$--$\omega$ mixing on the normalization can be neglected). 
However, this naive observation is incorrect. 
A similar case is the pion electromagnetic form factor. In a specific large-$N_c$ QCD model, dual-$\rm{QCD}_{\infty}$~\cite{Dominguez:2001zu}, Euler's beta function of the Veneziano type requires an infinite spectral tower of zero-width resonances, with masses and couplings following Regge trajectories. 
An important feature of this realization is that the coupling constants of the resonance tower alternate in sign~\cite{Dominguez:2001zu}. In other words, the contributions of $\rho$ and $\rho^{\prime}$ have opposite signs, which is verified in real $\rm{QCD}_3$ by $e^{+} e^{-} \to \pi^{+} \pi^{-}$ data~\cite{BaBar:2012bdw}.
Similarly, the opposite signs for the $\rho$ and $\rho^{\prime}$ contributions to the $J / \psi \to \pi^0 \gamma^\ast$ TFF is also consistent with the dual-$\rm{QCD}_{\infty}$ model. 
In Sec.~\ref{sec:pQCD} we will present a more robust argument based on the superconvergence sum rule~\eqref{eq:sumrule_v3}. Therefore, the contributions from the $2\pi$ intermediate states and the $\rho^{\prime}$ roughly contribute about $80\%-20\%=60\%$ of the sum rule for normalization of TFF. 
The remaining portion comes from higher resonances, such as charmonium(-like) states.\footnote{One might be concerned about the neglect of the $\rho^{\prime \prime}(1700)$ contribution. It is difficult to quantify this contribution due to the lack of experimental data. Given current data, omitting contributions from $\rho^{\prime \prime}(1700)$ and higher excited states remains a reasonable approximation.}

\subsection{Charmonium contributions}

As for the charmonium contributions, the TFF can be parameterized using a simple monopole ansatz, also referred to as ``VMD"~\cite{Fu:2011yy},
\begin{align}\label{eq:monopole}
    f_{\psi \pi^0}^{(c \bar{c})}(s)=\frac{f_{\psi \pi^0}^{(c \bar{c})}(0)}{1-s / \Lambda^2},
\end{align}
where $f_{\psi \pi^0}^{(c \bar{c})}(0)=\left|f_{\psi \pi^0}^{(c \bar{c})}(0)\right|e^{i \delta^{c \bar{c}}}$ is a complex number. The pole mass $\Lambda$ is fixed to the lowest $1^{--}$ charmonium mass $J/\psi$, since this is the lowest vector charmonium, rather than that of $\psi(2S)$ used in Ref.~\cite{Fu:2011yy}. 
Note that this ansatz is less reliable than VMD in the light-quark sector, since the mass splittings between various vector charmonium(-like) states are small compared to the difference between the charmonium mass and the $s$ region of interest.

The modulus $\left|f_{\psi \pi^0}^{(c \bar{c})}(0)\right|$ and the phase $\delta^{c \bar{c}}$ are not independent if we further require that the overall form factor normalization in Eq.~\eqref{eq:normalization_psipi0} is saturated by $2\pi$, $\rho'$ and charmonium contributions, 
\begin{align}\label{eq:constraint}
    \left|f_{\psi \pi^0}^{(2\pi)}(0)+f_{\psi \pi^0}^{(\rho^\prime)}(0)+f_{\psi \pi^0}^{(c \bar{c})}(0) \right|=\left|f_{\psi \pi^0}(0)\right|.
\end{align}
Here, we assign the same sign to $f_{\psi \pi^0}^{(2\pi)}(0)$ and $f_{\psi \pi^0}^{(c\bar c)}(0)$, while explicitly introducing an opposite sign for the $\rho^\prime$ contribution [cf.\ Eq.~\eqref{eq:constraint}] for the reason discussed above. This corresponds to choosing the sign convention such that $\operatorname{sgn}\left(g_{J / \psi \rho^{\prime} \pi} g_{\rho^{\prime} \gamma}\right)=-1$. From Eq.~\eqref{eq:constraint}, we obtain
\begin{align}\label{eq:fcc_modulus}
    \left|f_{\psi \pi^0}^{(c \bar{c})}(0)\right|=-f^{(2\pi,\rho^\prime)}_{\psi \pi^0}(0)\cos\delta^{c\bar c}+\sqrt{\left|f_{\psi \pi^0}(0)\right|^2-\left(f^{(2\pi,\rho^\prime)}_{\psi \pi^0}(0)\right)^2 \sin^2\delta^{c\bar c}},
\end{align}
where $f^{(2\pi,\rho^\prime)}_{\psi \pi^0}(0)\equiv f_{\psi \pi^0}^{(2\pi)}(0)+f_{\psi \pi^0}^{(\rho^\prime)}(0)$.

\section{Asymptotic behavior and sum rules}\label{sec:pQCD}

The pQCD asymptotic behavior requires the TFF to vanish as $f_{\psi\pi} \sim 1/s^2$ as $s \to \infty$. 
Note that we have two types of contributions here: one from light quark degrees of freedom, including $2\pi$, $3\pi$ ($\omega$ and $\phi$), and $4\pi$ ($\rho^{\prime}$); and the other from heavy-quark degrees of freedom, specifically the charmonium $c\bar{c}$ contribution. In principle, both types should satisfy the asymptotic behavior separately.
The TFF in this work is constructed from precise low-energy phenomenology, without directly matching to pQCD in the high-$s$ region. Nevertheless, the pQCD asymptotics implicitly impose certain constraints on relative signs of contributions in the same sector and limit the applicability of a single monopole ansatz in the high-energy region.

For the light-quark contribution in the limit $s\to \infty$, one can read off a $\mathcal{O}(s^0)$ sum rule from Eq.~\eqref{eq:2pi3pi_DR} as
\begin{align}\label{eq:sumrule_v2}
    0= &\, f_{\psi \pi^0}^{(2\pi)}(0)-\frac{1-\tilde{\epsilon}_{\rho \omega}}{96 \pi^2} \int_{4 M_\pi^2}^{\infty} \mathrm{d} s^{\prime} \sigma_\pi^3\left(s^{\prime}\right)F_\pi^{V *}\left(s^{\prime}\right)f_1\left(s^{\prime}\right) \nonumber\\
    & -w_{\psi \omega \pi} \left[1-\frac{\tilde{\epsilon}_{\rho \omega}}{48 \pi^2 g_{\omega \gamma}^2} \int_{4 M_\pi^2}^{\infty} \mathrm{d} s^{\prime} \frac{\sigma_\pi^3\left(s^{\prime}\right)\left|F_\pi^V\left(s^{\prime}\right)\right|^2}{s^{\prime}}\right]-w_{\psi \phi \pi}.
\end{align}
Note that 
\begin{align}\label{eq:sumrule_v4}
    f_{\psi \pi^0}^{(2\pi)}(0)=\frac{1}{96 \pi^2} \int_{4 M_\pi^2}^{\infty} \mathrm{d} s^{\prime} \sigma_\pi^3\left(s^{\prime}\right)F_\pi^{V *}\left(s^{\prime}\right)f_1\left(s^{\prime}\right)
\end{align}
is simply the sum rule for normalization and $\tilde{\epsilon}_{\rho \omega} \ll 1$. The saturation of this sum rule can be checked by the magnitude of the right-hand side of Eq.~\eqref{eq:sumrule_v4}, which equals $-3.84\times 10^{-5}\,\mathrm{GeV}^{-1}$ or $-4.22\times10^{-5} \,\mathrm{GeV}^{-1}$, corresponding to the two values of $w_{\psi \phi \pi}$ in Eq.~\eqref{eq:modulus_w},\footnote{Here, we assume that $w_{\psi \omega \pi}=\left|w_{\psi \omega \pi}\right|$ and $w_{\psi \phi \pi}=-\left|w_{\psi \phi \pi}\right|$ for the moment.} respectively. These nonzero contributions arise almost entirely from the isovector vector pole. It follows that the contributions of the isovector vector poles in the sum rule~\eqref{eq:sumrule_v2} result in a less than $10\%$ violation of the normalization sum rule~\eqref{eq:sumrule_v4}; therefore the latter is a very good approximation to the former.
From the $\mathcal{O}(s^{-1})$ terms of Eq.~\eqref{eq:2pi3pi_DR} and Eq.~\eqref{eq:rhoprime_DR}, one can derive another superconvergence sum rule as 
\begin{align}\label{eq:sumrule_v3}
    0= & -\frac{1-\tilde{\epsilon}_{\rho \omega}}{96 \pi^2} \int_{4 M_\pi^2}^{\infty} \mathrm{d} s^{\prime}~s^\prime\sigma_\pi^3\left(s^{\prime}\right)F_\pi^{V *}\left(s^{\prime}\right)f_1\left(s^{\prime}\right) \notag\\
    & +\frac{\tilde{\epsilon}_{\rho \omega}}{96 \pi^2}\left(M_\omega^2-iM_\omega\Gamma_\omega\right) \int_{4 M_\pi^2}^{\infty} \mathrm{d} s^{\prime}~\sigma_\pi^3\left(s^{\prime}\right)F_\pi^{V *}\left(s^{\prime}\right)f_1\left(s^{\prime}\right) \nonumber\\
    & - w_{\psi \omega \pi} \left(M_\omega^2-iM_\omega\Gamma_\omega\right) \left[1-\frac{\tilde{\epsilon}_{\rho \omega}}{48 \pi^2 g_{\omega \gamma}^2} \int_{4 M_\pi^2}^{\infty} \mathrm{d} s^{\prime} \frac{\sigma_\pi^3\left(s^{\prime}\right)\left|F_\pi^V\left(s^{\prime}\right)\right|^2}{s^{\prime}}\right] \nonumber\\
    & +w_{\psi \omega \pi} \frac{\tilde{\epsilon}_{\rho \omega}}{48 \pi^2 g_{\omega \gamma}^2} \int_{4 M_\pi^2}^{\infty} \mathrm{d} s^{\prime} \frac{\sigma_\pi^3\left(s^{\prime}\right)\left|F_\pi^V\left(s^{\prime}\right)\right|^2}{s^{\prime}}-w_{\psi \phi \pi} \left(M_\phi^2-iM_\phi\Gamma_\phi\right) \nonumber\\
    & -\frac{1}{\pi}\int_{4 M_\pi^2}^{\infty} \mathrm{d} s^{\prime}~\frac{g_{J/\psi \rho^\prime \pi} g_{\rho^\prime\gamma} \sqrt{s^\prime} \Gamma_{\rho^{\prime}}(s^\prime)}{\left(M_{\rho^{\prime}}^2-s^\prime\right)^2+s^\prime \Gamma_{\rho^{\prime}}^2(s^\prime)},
\end{align}
where we choose $\operatorname{sgn}\left(g_{J/\psi \rho^\prime \pi} g_{\rho^\prime\gamma} \right)=-1$ to match the convention of Eq.~\eqref{eq:constraint}. In fact, the first and last terms of Eq.~\eqref{eq:sumrule_v3} dominate this sum rule, resulting in $(-3.50+0.11i)\times 10^{-4}\,\mathrm{GeV}+2.61\times 10^{-4}\,\mathrm{GeV}=(-0.89+0.11i)10^{-4}\,\mathrm{GeV}$. Note that the first integral is in fact logarithmically divergent. To obtain a reasonable value we therefore impose an integral cutoff $\Lambda=(2~\text{GeV})^2$ on both the first and the last integrals. This cancellation at the same time reduces the sensitivity to the cutoff. This is motivated by the fact that both the $2\pi$ contribution obtained from the dispersion representation and the $4\pi$ contribution associated with $\rho'$ are restricted to low energies and hence are not valid in the pQCD asymptotic region. The discontinuities of these contributions (and of further channels) necessitate a fine-tuned cancellation as $s\to\infty$ so as to produce the $1/s^2$ scaling behaviour, which ensures that the sum of the two integrals and the superconvergence sum rule converges. This cancellation follows from pQCD asymptotics. It also shows that the contributions of $2\pi$ and $\rho^\prime$ should have opposite signs. This behavior is exactly what one expects from large-$N_c$ considerations, and finds phenomenologically for light-vector-meson transition form factors such as $\omega\to\pi^0\gamma^*$~\cite{Achasov:2016zvn} and $\rho\to\eta\gamma^*$~\cite{BaBar:2007qju,BaBar:2018erh,Holz:2015tcg}.

However, for the heavy-quark case, a single monopole in Eq.~\eqref{eq:monopole} does not satisfy the pQCD constraint. If more poles are introduced, their contributions must be carefully balanced to fulfill the high-energy behavior required by pQCD. 
In practice, we prefer to use a minimal number of parameters and determine the TFF using precise low-energy empirical inputs, thereby relaxing the pQCD constraint for the charmonium sector.
We also present an alternative parameterization using a dipole form in Appendix~\ref{app:dipole}, although this approach requires introducing at least one additional free parameter to the fit.

\section{Results and Discussion}\label{sec:Results}

\subsection{Fit results}

The total TFF is given by
\begin{align}
    f_{\psi \pi^0}(s)=f_{\psi \pi^0}^{(2\pi,3\pi)}(s)+f_{\psi \pi^0}^{(\rho^{\prime})}(s)+f_{\psi \pi^0}^{(c \bar{c})}(s) . 
\end{align}
We fix $f_{\psi \pi^0}^{(2 \pi)}(0)$ in Eq.~\eqref{eq:2pi3pi_DR} using the sum rule in Eq.~\eqref{eq:sumrule_v4} [sum rule~\eqref{eq:sumrule_v2} yields similar values]. This approximately ensures that the contribution from light degrees of freedom satisfies the pQCD asymptotic constraint. 
For the coupling constants $g_{\rho^\prime \gamma}$ and $g_{J/\psi\rho^\prime \pi}$ related to the $\rho^{\prime}(1450)$, we adopt the results from the phenomenological analysis, 
\begin{align}\label{eq:rho'coupling}
    \left|g_{\rho^{\prime} \gamma}\right|=0.16 ~\mathrm{GeV}^2\ ,\quad \left|g_{J / \psi \rho^{\prime} \pi}\right|=2.73_{-0.80}^{+0.61} \times 10^{-3} ~\mathrm{GeV}^{-1}.
\end{align}
Relaxing the ranges of these coupling constants leads to unstable fits, since there is only a single data point above 1.3~GeV with large errors. 
Moreover, we use the values for $\left|w_{\psi\omega\pi}\right|$ and $\left|w_{\psi\phi\pi}\right|$ from the pole dominance ansatz [cf.\ Eq.~\eqref{eq:modulus_w}].
Thus, a priori we are left with three real parameters: the phases of $w_{\psi\omega\pi}$ and $w_{\psi\phi\pi}$, and the phase $\delta^{c\bar c}$.
Since there are at most two experimental data points in the energy region near the $\phi$, we fix the phase of $w_{\psi\phi\pi}$ such that $w_{\psi \phi \pi}=-\left|w_{\psi \phi \pi}\right|$.
Thus we are left with only two free parameters.

The BESIII Collaboration reported two distinct solutions for the $J/\psi\to \phi \pi^0$ branching ratio, $\mathcal{BR}(J/\psi\to \phi \pi^0)=3\times 10^{-6}$ or $1\times 10^{-7}$, due to the multisolution ambiguity arising from different interference patterns ($J/\psi \rightarrow \phi \pi^0$ and $J/\psi \rightarrow K^{+} K^{-} \pi^0$) in experimental analyses~\cite{BESIII:2015asx}. 
The larger (smaller) branching ratio corresponds to the larger (smaller) weight factor in Eq.~\eqref{eq:modulus_w}.
Since we cannot distinguish between the two scenarios at present, we implement two fits: Theory-I corresponds to the larger branching ratio, whereas Theory-II corresponds to the smaller one.

To fit the TFF, we define the $\chi^2$ function as
\begin{align}
    \chi^2=\sum_i\left(\frac{\left|F_{\psi \pi^0}\left(s_i\right)\right|_{\mathrm{exp}}^2-\left|F_{\psi \pi^0}\left(s_i\right)\right|_{\mathrm{th}}^2}{\sigma_i}\right)^2,
\end{align}
where $F_{\psi \pi^0}\left(s\right)$ is the normalized TFF [cf.\ Eq.~\eqref{eq:normalized TFF}]. $\left|F_{\psi \pi^0}(s)\right|_{\exp }$ and $\sigma_i$ are the experimental normalized TFF and its uncertainty, respectively, obtained from Ref.~\cite{BESIII:2025xjh}. $\left|F_{\psi \pi^0}(s)\right|_{\text {th }}^2$ is the average in each bin $[\sqrt{s_i}-\Delta_i, \sqrt{s_i}+\Delta_i]$ of our normalized TFF:
\begin{align}
    \left|F_{\psi \pi^0}\left(s_i\right)\right|_{\text {th}}^2=\frac{1}{4 \sqrt{s_i} \Delta_i} \int_{\left(\sqrt{s_i}-\Delta_i\right)^2}^{\left(\sqrt{s_i}+\Delta_i\right)^2} \md s\left|F_{\psi \pi^0}(s)\right|^2,
\end{align}
where $\Delta_i$ represents half of the bin width. To minimize the $\chi^2$ function, we fit the data using MINUIT~\cite{James:1975dr,iminuit}. 
The parameter values and the correlation coefficient from two fits are listed in Table~\ref{tab:Fit}. 
Additionally, to estimate theoretical uncertainties beyond those from the fitting procedure, we perform multiple fits with different inputs, including: varying the TFF normalization within experimental constraints, namely, $\left|f_{\psi \pi^0}(0)\right|=6.0(3) \times 10^{-4} \mathrm{GeV}^{-1}$; varying the $\pi\pi$ phase shift input by using solution~2 shown in Fig.~\ref{fig:phase}; varying the modulus of two coupling magnitudes within the ranges of Eq.~\eqref{eq:modulus_w}; varying the coupling constants corresponding to $\rho^{\prime}(1450)$ within the ranges of Eq.~\eqref{eq:rho'coupling}; and changing the energy-dependent width of $\rho^{\prime}(1450)$ as discussed in Sec.~\ref{sec:rho'}. 
All uncertainties from the above variations are added in quadrature to the final TFF results.

\begin{table}[t]
    \centering
    \caption{Parameter values and correlation coefficient from two fits.}
\begin{tabular}{lcc}
\hline \hline
Parameters & Theory-I & Theory-II \\
\hline 
$\delta_{\psi\omega\pi}$ & $0.93(26)$ & $0.88(27)$ \\
$\delta^{c\bar c}$ & $0.90(20)$ & $0.82(21)$  \\
\hline 
$\chi^2 /$ d.o.f. & ~$28.20/(30-2)=1.01$~ & ~$31.09/(30-2)=1.11$ \\
Correlation coefficient & $0.70$ & $0.71$ \\
\hline \hline
\end{tabular}
    \label{tab:Fit}
\end{table}
From Table~\ref{tab:Fit}, as for $J/\psi\to\phi\pi^0$, our fitting result slightly favors the solution with the larger branching ratio, although the solution with the smaller one cannot be ruled out.
Unless otherwise specified, all the following calculations are based on the results from Theory-I.

\begin{figure}[tb]
    \centering
    \includegraphics[width=1\linewidth]{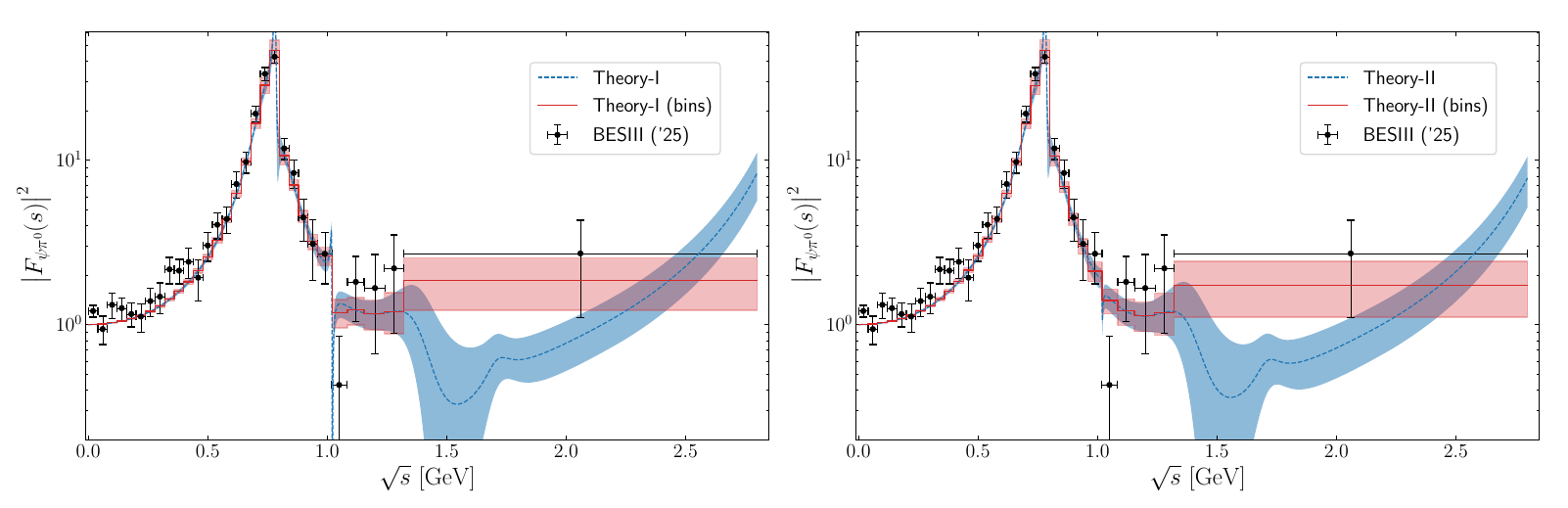}
    \caption{Normalized $J/\psi\to\pi^0 \gamma^\ast$ TFF as a function of energy $\sqrt{s}$ (in logarithmic scale). \textit{Left:} fit Theory-I; \textit{right:} fit Theory-II. The black points represent the experimental data from BESIII~\cite{BESIII:2025xjh}, while the red solid line represents the result of our fit averaged in each bin, with the parameters given in Table~\ref{tab:Fit}. The blue dashed line represents the same TFF but computed directly as a continuous function of $\sqrt{s}$.}
    \label{fig:Fit}
\end{figure}
\begin{figure}[t]
    \centering
    \includegraphics[width=.8\linewidth]{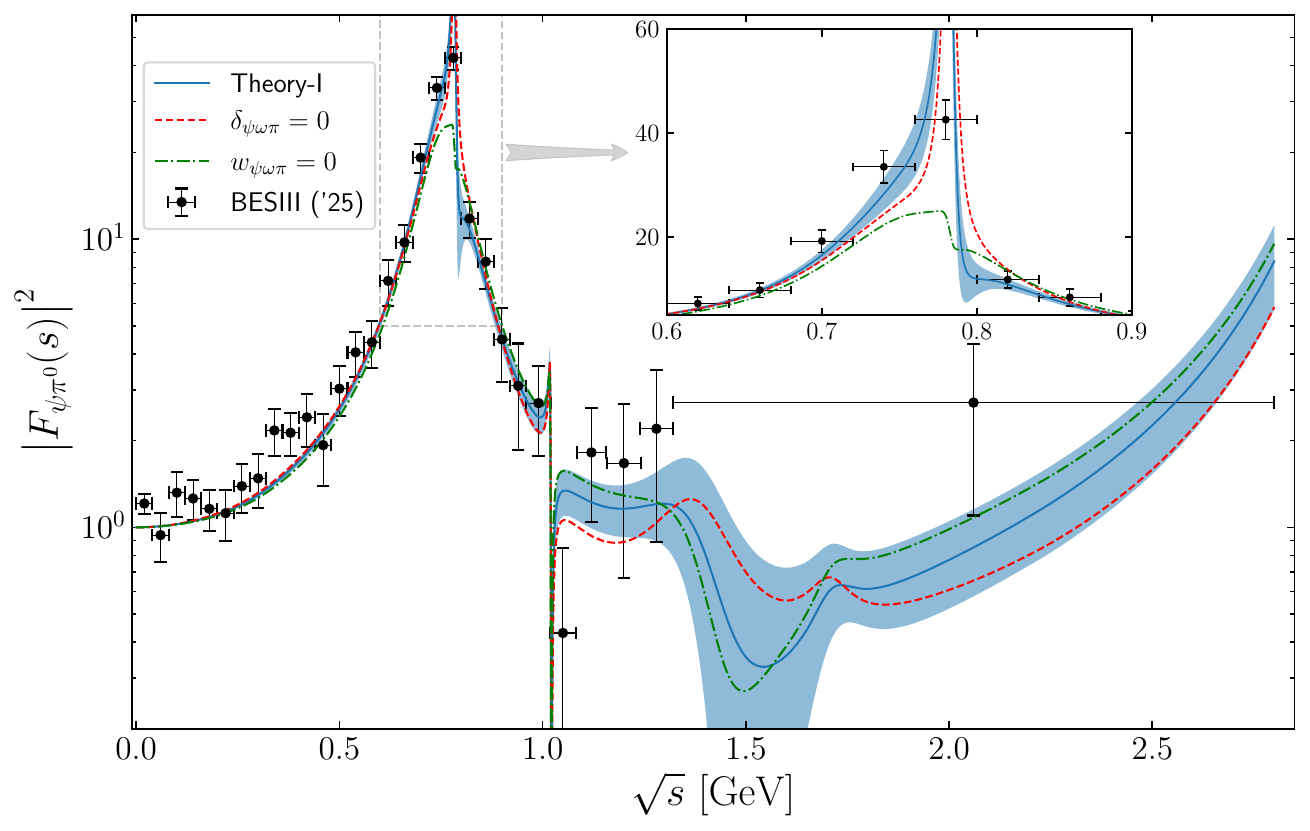}
    \caption{Same as Fig.~\ref{fig:Fit}, except that the red dashed (green dot-dashed) curves correspond to the one-parameter fit with $\delta_{\psi\omega\pi}=0$ ($w_{\psi\omega\pi}=0$).}
    \label{fig:Omega=0}
\end{figure}

In Fig.~\ref{fig:Fit}, we show the squared absolute value of the normalized TFF from our fits. As can be seen, our result shows excellent agreement with the experimental data from BESIII~\cite{BESIII:2025xjh} across the full energy range from the $\pi\pi$ threshold to about 2.8~GeV, which extends far beyond the $\rho$ pole regime. Remarkably, our analysis uses only two parameters, which is a significant improvement compared to the seven-parameter fit adopted in the BESIII analysis~\cite{BESIII:2025xjh}.
An important observation is the presence of a significant structure, with an abrupt drop at the $\omega$ pole for the $J/\psi\to\pi^0 \gamma^\ast$ TFF. 
To isolate the $\omega$ contribution, we perform a simplified fit with $w_{\psi\omega\pi}$ fixed to zero. The resulting $\chi^2/\text{d.o.f}=77.40/(30-1)=2.67$ is far worse, and the corresponding curve is the green dot-dashed line in Fig.~\ref{fig:Omega=0} (see also the result in Ref.~\cite{Kubis:2014gka}). The result clearly indicates that neglecting the isospin-violating $\omega$ pole term in Eq.~\eqref{eq:2pi3pi_DR} ($\propto w_{\psi\omega\pi}$) fails to describe the data, in particular the abrupt drop at the location of the $\omega$ mass.
This effect has not been observed in experimental studies of the $\phi \to \pi^0 \gamma^\ast$ process~\cite{KLOE-2:2016pnx} (naturally, due to kinematic constraints, $\omega \to$ $\pi^0 \gamma^\ast$ cannot exhibit the $\rho$--$\omega$ mixing structure). 
Since the processes $J/\psi,\phi \to \omega \pi^0$ violate isospin, previous theoretical analyses~\cite{Schneider:2012ez,Kubis:2014gka,Danilkin:2014cra,JPAC:2020umo} did not consider $\rho$--$\omega$ mixing, or simply added an extra BW form to the $J/\psi, \phi$ decay amplitudes for the $\omega$ pole to simulate its contribution~\cite{Garcia-Lorenzo:2025uzc}. 
Interestingly, all these predictions are slightly lower than the experimental values near the $\rho$ pole (for a comparison, see Fig.~3 of Ref.~\cite{BESIII:2025xjh}). According to our results, this strongly suggests that the isospin-violating $\omega$ pole and $\rho$--$\omega$ mixing contribute significantly to resolving this discrepancy. 
Our dispersive framework, which includes $\rho$--$\omega$ mixing effects in a self-consistent manner, can provide a benchmark for future dispersive analyses of the TFF. 
Moreover, our analysis reveals a nontrivial structure in the $\rho^{\prime}(1450)$ region within $1.3\sim1.6$~GeV, suggesting the need for more precise experimental measurements in this energy range.

\begin{figure}[t]
    \centering
    \includegraphics[width=1\linewidth]{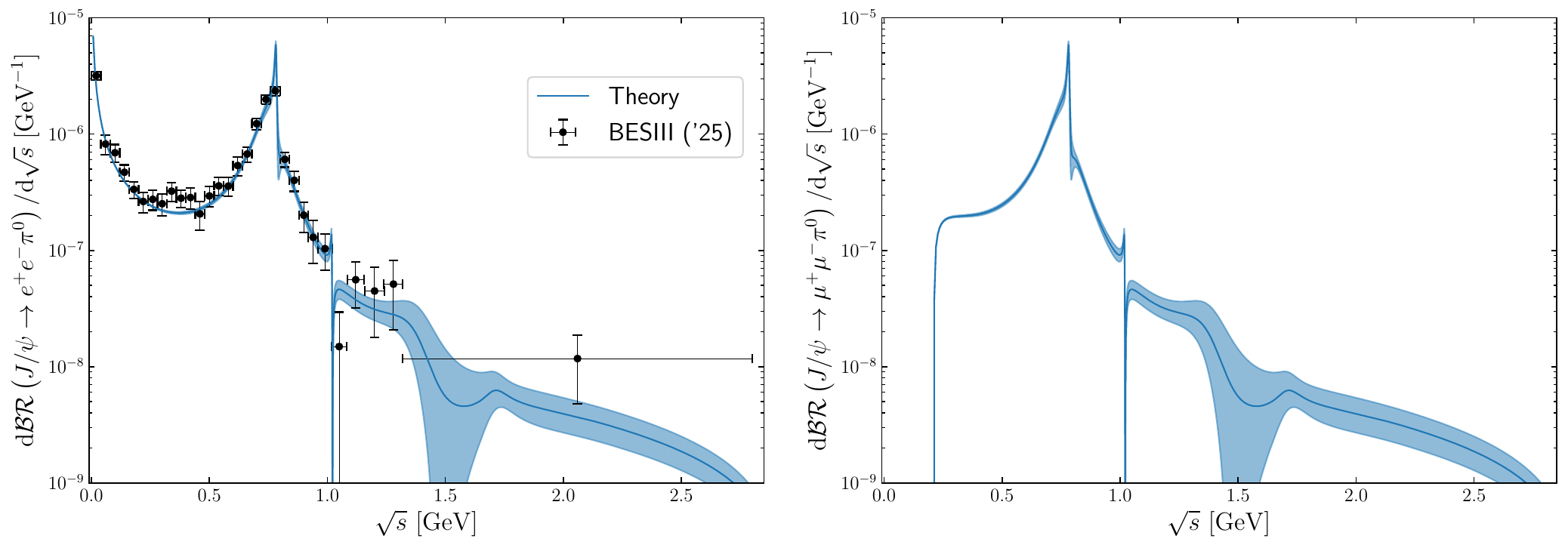}
    \caption{Differential branching fractions $\mathrm{d} \mathcal{BR} / \mathrm{d} \sqrt{s}$ for $J / \psi \to \pi^0 e^{+} e^{-}$ (left) and $J / \psi \to \pi^0 \mu^{+} \mu^{-}$ (right).}
    \label{fig:diffBR}
\end{figure}

In addition, using the best fit Theory-I we present in Fig.~\ref{fig:diffBR} the differential branching fractions for both $J / \psi \to \pi^0 e^{+} e^{-}$ and $J / \psi \to \pi^0 \mu^{+} \mu^{-}$ using Eq.~\eqref{eq:diff_Gamma}. Both channels exhibit prominent features: a pronounced $\rho$ resonance dominating the TFF structure, a nontrivial structure around the $\phi$ pole, which would be much less pronounced when employing Theory-II, and a strong low-energy enhancement for $J / \psi \to \pi^0 e^{+} e^{-}$ near $\sqrt{s}=0$ arising from the photon pole.
By integrating over the respective spectra, we obtain the branching fractions for the two dilepton final states as
\begin{align}
    \mathcal{BR}\left(J / \psi \rightarrow \pi^0 e^{+} e^{-}\right)=7.03^{+0.62}_{-0.60} \times 10^{-7}, \label{eq:BR_e}\\
    \mathcal{BR}\left(J / \psi \rightarrow \pi^0 \mu^{+} \mu^{-}\right)=4.15^{+0.58}_{-0.50} \times 10^{-7} .
\end{align}

\begin{table}[t]
    \centering
    \caption{Branching ratios $\left(\times 10^{-7}\right)$ for $J / \psi \rightarrow \pi^0 \ell^{+} \ell^{-}$, where $\ell=e,\mu$.}
\begin{tabular}{lcccccc}
\hline\hline
& Experiment~\cite{BESIII:2025xjh} & This Work & DR~\cite{Kubis:2014gka} & RChT~\cite{Yan:2023nqz} & RChT~\cite{Chen:2014yta} & VMD~\cite{Fu:2011yy}  \\
\hline $J/\psi \rightarrow \pi^0 e^{+} e^{-}$ & $8.06 \pm 0.31(\text{stat}) \pm 0.38(\text{syst})$ & $7.03^{+0.62}_{-0.60}$ & $(5.5 \ldots 6.4)$ & $12.94 \pm 0.44$ & $11.91 \pm 1.38$ & $3.89_{-0.33}^{+0.37}$ \\
$J/\psi \rightarrow \pi^0 \mu^{+} \mu^{-}$ & --- & $4.15^{+0.58}_{-0.50}$ & $(2.7 \ldots 3.3)$ & $3.04 \pm 0.10$ & $2.80 \pm 0.32$ & $1.01_{-0.09}^{+0.10}$ \\
\hline\hline
\end{tabular}
    \label{tab:BR}
\end{table}
The value in Eq.~\eqref{eq:BR_e} can be compared with previous dispersive analyses~\cite{Kubis:2014gka} that did not consider $\rho$--$\omega$ mixing, as shown in Table~\ref{tab:BR}.
This result is also in good agreement with the value extracted by BESIII within uncertainties~\cite{BESIII:2025xjh}. A comparison with the predictions of VMD~\cite{Fu:2011yy} and resonance chiral theory (RChT)~\cite{Chen:2014yta,Yan:2023nqz} is also shown in Table~\ref{tab:BR}.

\subsection{Phase between strong and electromagnetic amplitudes}

There are qualitatively two kinds of contributions to the hadronic decays of $J/\psi$ into $\rho^0\pi^0$ and $\omega\pi^0$, i.e., via strong interaction (e.g., three gluons or charmed-meson loops) and via one virtual photon.
While isospin symmetry is preserved to a very good approximation in the former decay, it is broken in the latter.
The one-photon electromagnetic current can be decomposed as $\frac{2}{3} \bar{u} \gamma_\mu u-\frac{1}{3} \bar{d} \gamma_\mu d=\frac{1}{6} J_\mu^s+\frac{1}{2} J_\mu^v$ with $J_\mu^{s,v}=\bar{u} \gamma_\mu u\pm \bar{d} \gamma_\mu d$ representing isoscalar and isovector vector currents, respectively. 
From this decomposition, one concludes that the one-photon contribution to the isoscalar $\rho^0\pi^0$ is three times  smaller than that to the isovector $\omega\pi^0$. 
Usually, the isospin breaking through strong interactions, $\sim \mathcal{O}((m_d-m_u)/\Lambda_{\rm QCD})$, is of the same order as that from virtual photons, $\sim \mathcal{O}(\alpha)$.
The branching fraction of $J/\psi\to \rho^0\pi^0$, $6.2(6) \times 10^{-3}$, is one order of magnitude larger than the branching fraction of $J/\psi\to\omega\pi^0$, $4.5(5) \times 10^{-4}$~\cite{ParticleDataGroup:2024cfk}.
Hence, we expect that $J/\psi\to \rho^0\pi^0$ should proceed mainly through strong interaction, which is stronger than the one-photon contribution by roughly one order of magnitude at the amplitude level.
The strong-interaction contribution to the isospin-breaking $\omega\pi^0$ mode should be further suppressed compared with the $\rho^0\pi^0$ mode by $\mathcal{O}((m_d-m_u)/\Lambda_{\rm QCD})$, thus should be significantly smaller than the one-photon contribution to the $\omega\pi^0$ mode. The photon-dominance picture for a vector charmonium decay into $\omega\pi^0$ can be checked by comparing the following ratios of partial widths~\cite{ParticleDataGroup:2024cfk}:
\begin{align}
    \frac{\Gamma(\psi'\to\omega \pi^0)}{\Gamma(J/\psi\to\omega \pi^0)}&=\frac{293(9)~\rm{keV}\times (2.1(6)\times10^{-5})}{92.6(1.7)~\rm{keV}\times(4.5(5)\times10^{-4})}=0.148(46),\nonumber\\
    \frac{\Gamma(\psi'\to\gamma \pi^0)}{\Gamma(J/\psi\to\gamma \pi^0)} &=\frac{293(9)~\rm{keV}\times (1.04(22)\times10^{-6})}{92.6(1.7)~\rm{keV}\times(3.39(8)\times10^{-5})}=0.097(21).
\end{align}
The two ratios are approximately equal within uncertainties.
Therefore, we expect that the $\rho^0\pi^0$ and $\omega\pi^0$ modes should be dominated by the strong interaction and the one-photon mechanisms, respectively.

This can be verified in a more direct manner. As a first step, we approximate Eq.~\eqref{eq:2pi3pi_DR} by a VMD version of the TFF,
\begin{align}\label{eq:TFF_VMD}
    f^{(2\pi,3\pi),\rm{VMD}}_{\psi\pi^0}(s)= &\, f^{(2\pi)}_{\psi\pi^0}(0) \nonumber\\
    & +\frac{w_{\psi \rho \pi} s}{M_\rho^2-s-i M_\rho \Gamma_\rho}\left[1+\frac{\tilde{\epsilon}_{\rho \omega} s}{M_\omega^2-s-i M_\omega \Gamma_\omega}\right] \nonumber\\
    & +\frac{w_{\psi \omega \pi} s}{M_\omega^2-s-i M_\omega \Gamma_\omega}\left[1+\frac{9 \tilde{\epsilon}_{\rho \omega} s}{M_\rho^2-s-i M_\rho \Gamma_\rho}\right]+\ldots,
\end{align}
using the narrow width (NW) approximation~\cite{Holz:2022hwz} 
\begin{align}
    \frac{s}{48\pi^2}\int_{4 M_\pi^2}^{\infty} \mathrm{d} s^{\prime} \frac{\sigma_\pi^3\left(s^{\prime}\right)\left|F_\pi^V\left(s^{\prime}\right)\right|^2}{s^{\prime}\left(s^{\prime}-s-i \epsilon\right)}\overset{\rm{NW}}{\longrightarrow}\frac{g_{\rho\gamma}^2 s}{M_\rho^2-s-i M_\rho \Gamma_\rho},
\end{align}
and the VMD relation $g_{\rho\gamma}=3\rho_{\omega\gamma}$. Consequently, the issue reduces to determining the relation between the weight functions $w_{\psi \rho \pi}$ and $w_{\psi \omega \pi}$. Taking into account the isospin factor and the dominance of the $\omega\pi^0$ mode by the one‑photon mechanism, one finds for $w_{\psi\rho\pi}$ 
\begin{align}
    w_{\psi\rho\pi}\simeq w_{\psi\rho\pi}^{\rm{S}}+w_{\psi\rho\pi}^\gamma=w_{\psi\rho\pi}^{\rm{S}}+\frac{1}{3}w_{\psi\omega\pi}^\gamma\simeq w_{\psi\rho\pi}^{\rm{S}}+\frac{1}{3}w_{\psi\omega\pi},
\end{align}
with the superscripts ``$\rm{S}$'' and ``$\gamma$'' representing the contributions from the strong interaction and one virtual photon, respectively. By matching the BW parameterization of the $\rho$ meson ($\propto w_{\psi \rho \pi}$) in Eq.~\eqref{eq:TFF_VMD} with the corresponding dispersive integral in Eq.~\eqref{eq:2pi3pi_DR} and fitting over the energy range $\sqrt{s}=0.6-1.3~$GeV, we determine $w_{\psi \rho \pi}=4.76\,e^{-i\,0.13}\times10^{-4}\,\rm{GeV}^{-1}$, which compares to the value $\left|w_{\psi \rho \pi}\right|=5.36(27)\times10^{-4}\,\rm{GeV}^{-1}$ obtained using the NW approximation [cf.\ Eq.~\eqref{eq:modulus_w}]. 
We take the average $(4.76+5.36)/2 = 5.06$ as the central value of $\left|w_{\psi \rho \pi}\right|$ and assign a conservative $1\sigma$ uncertainty equal to half the difference, i.e., 0.3. Therefore, the value of the strong part, $w_{\psi\rho\pi}^{\rm{S}}$, is given by 
\begin{align}
    w_{\psi\rho\pi}^{\rm{S}}=w_{\psi\rho\pi}-\frac{1}{3}w_{\psi\omega\pi} &=\left[5.06(30)e^{-i\,0.13}\times 10^{-4}-1.46(8)e^{i\,0.93(26)}\times 10^{-5}\right]\rm{GeV}^{-1}\nonumber\\
    &=4.99(30)e^{-i\,0.16(10)}\times 10^{-4}\,\rm{GeV}^{-1}.
\end{align}
This result is compatible with our expectations. The $\rho\pi^0$ mode is dominated by the strong interaction, whereas the contribution of one virtual photon is of an order of magnitude smaller and affects the phase of $w_{\psi\rho\pi}^{\rm{S}}$ by no more than $2^\circ$.

To check the role played by the mentioned phase, we also present a simplified fit with $\delta_{\psi\omega\pi}$ fixed to zero.
As shown in Fig.~\ref{fig:Omega=0}, an a priori assumption of $\delta_{\psi\omega\pi}=0$ results in a slightly worse $\chi^2/\text{d.o.f}=36.75/(30-1)=1.27$ compared to the best value of 1.01 in Theory-I.
From comparing the red dashed line with the blue band, one sees that around the $\omega$ mass the case with the nonvanishing phase has a large slope on the left shoulder, while a smaller slope on the right shoulder. This means that $\rho$--$\omega$ mixing and its interplay with the $\omega$ pole contribution and, in particular, its phase relative to the $\rho$ contribution that is fully controlled within the dispersive scheme employed here, play a crucial role in forming the specific line shape. Thus the phase between strong interaction and one-photon contributions in $J/\psi$ hadronic decays can be extracted as $(62\pm 21)^\circ$, where the small phase of $w_{\psi\rho\pi}^{\rm{S}}$ is taken into account [omitting this correction would reduce the phase to 
$(53\pm 15)^\circ$].
Our result is consistent with the value $(72\pm 17)^{\circ}$~\cite{LopezCastro:1994xw} of the relative strong--electromagnetic phase from the global fit to $J/\psi \to V P$ ($P$, $V$ stand for light $0^{-}$and $1^{-}$ mesons) within uncertainties, while it deviates slightly (1.5$\sigma$) from $90^\circ$ corresponding to the absence of an interference pattern. It should be emphasized that while the relative phase near $90^\circ$ has been favored in phenomenological analyses~\cite{Gerard:1999uf}, there is no QCD foundation for this assumption. 
To improve the accuracy of our extraction, data with improved binning are necessary to define the line shape better.

\section{Summary and Outlook}\label{sec:Summary}

In this work, we have reanalyzed the $J / \psi \to \pi^0 \gamma^\ast$ transition form factor using the Khuri--Treiman framework, paying particular attention to the contributions of isospin-violating $\omega$ and $\phi$ intermediate states and their mixing effects with the two-pion ($\rho$) channel within the dispersive approach. 
This framework incorporates crossing symmetry while maintaining unitarity and analyticity. Above 1~GeV, contributions from the $\rho^{\prime}(1450)$ and the isospin-violating charmonium contribution have been included to account for the high-energy tail. 
The BESIII data for the transition form factor in the energy range from 0 to 2.8~GeV are very well described using only two fitting parameters---all other parameters were fixed from phenomenological input provided from other reactions. In particular, we predict a possible nontrivial structure around 1~GeV due to the $\phi$ contribution and a dip structure in the 1.2 to 1.6~GeV region caused by the $\rho^{\prime}$. 
We note that the prominence of the structure around the $\phi$ pole is governed by the experimentally measurable branching ratio of $J/\psi \to \phi \pi^0$. Compared to a previous dispersive analysis of $J / \psi \to \pi^0 e^{+} e^{-}$, we find that $\rho$--$\omega$ mixing and its interplay with the $\omega$ pole contribution lead to a non-negligible increase in the branching ratio.
Furthermore, this analysis allows us to extract the relative phase between the strong and one-virtual-photon (electromagnetic) modes in $J/\psi \to \rho \pi^0$ to be $(62 \pm 21)^{\circ}$. The result minimizes model dependence to the greatest extent possible, advancing our understanding of the long-standing  mysterious $\rho \pi$ puzzle in $J/\psi$ decays.

Intriguingly, the Khuri--Treiman framework applied to $\omega, \phi \to 3\pi$ decays produces theoretical results that fall slightly below experimental measurements in the 600--700~MeV range~\cite{Schneider:2012ez,JPAC:2020umo, Garcia-Lorenzo:2025uzc}. 
Thus, an accurate description must include isospin-violating contributions, with $\rho$--$\omega$ mixing as an important component.
We hope that our reevaluation will motivate more precise measurements of $J / \psi \to \pi^0 e^{+} e^{-}$ to constrain the $J / \psi \to \pi^0 \gamma^*$ transition form factor in the region beyond the $\rho$-peak and to further establish the phenomenology of the transition form factor advocated here.

\section*{Note Added}

After this work appeared on arXiv, a related study~\cite{Zhang:2025vbf} was finished on the $J/\psi\to\pi^0\gamma^*$ transition form factor within resonance chiral theory. That work also finds that $J/\psi \to \pi^0\rho^0$ is dominated by the strong interaction, whereas $J/\psi \to \pi^0\omega$ and $J/\psi \to \pi^0\phi$ are driven primarily by electromagnetic transitions, in qualitative agreement with our findings.

\begin{acknowledgements}

This work is supported in part by National Natural Science Foundation of China under Grants No.~12125507, No.~12361141819, and No.~12447101; by the Chinese Academy of Sciences under Grant No.~YSBR-101; and by the Postdoctoral Fellowship Program of China Postdoctoral Science Foundation under Grants No.~GZC20232773 and No.~2023M74360. C.H. also thanks the CAS President's International Fellowship Initiative (PIFI) under Grant No.~2025PD0087 for partial support. 

\end{acknowledgements}

\appendix
\section{Dipole form}\label{app:dipole}

A dipole parameterization can be simply formed by adding two single-pole terms,
\begin{align}
    f_{\psi \pi^0}^{(c \bar{c})}(s)=\frac{\tilde{f}_{\psi \pi^0}^{(c \bar{c})}(0)}{1-s / \Lambda_1^2}+\frac{\tilde{f}_{\psi \pi^0}^{(c \bar{c})\prime}(0)}{1-s / \Lambda_2^2}.
\end{align}
The effective zero-width pole masses $\Lambda_{1,2}$ should correspond to the masses of $1^{--}$ resonances near the energy scale of the decaying particle. Here, we fix $\Lambda_1=M_{J/\psi}$ and treat $\Lambda_2$ as a free parameter. The vanishing of the $\mathcal{O}(s^{-1})$ contribution enforced by the pQCD asymptotic constraint implies
\begin{align}
    \tilde{f}_{\psi \pi^0}^{(c \bar{c})\prime}(0)=-\tilde{f}_{\psi \pi^0}^{(c \bar{c})}(0)\frac{\Lambda_1^2}{\Lambda_2^2},
\end{align}
which results in
\begin{align}\label{eq:dipole}
    f_{\psi \pi^0}^{(c \bar{c})}(s)=\tilde{f}_{\psi \pi^0}^{(c \bar{c})}(0)\frac{\Lambda_1^2\left(\Lambda_2^2-\Lambda_1^2\right)}{\left(s-\Lambda_1^2\right)\left(s-\Lambda_2^2\right)}.
\end{align}
Similarly, if we define $\tilde{f}_{\psi \pi^0}^{(c \bar{c})}(0)=\left|\tilde{f}_{\psi \pi^0}^{(c \bar{c})}(0)\right| e^{i \delta^{c \bar{c}}}$, we can also derive an analog of Eq.~\eqref{eq:fcc_modulus},
\begin{align}\label{eq:fcc_modulud_v2}
    \left|\tilde{f}_{\psi \pi^0}^{(c \bar{c})}(0)\right|=\frac{\Lambda_2^2}{\Lambda_2^2-\Lambda_1^2}\left(-f^{(2\pi,\rho^\prime)}_{\psi \pi^0}(0)\cos\delta^{c\bar c}+\sqrt{\left|f_{\psi \pi^0}(0)\right|^2-\left(f^{(2\pi,\rho^\prime)}_{\psi \pi^0}(0)\right)^2 \sin^2\delta^{c\bar c}}\right).
\end{align}
As observed in Ref.~\cite{Kubis:2014gka}, we also find that strong cancellation effects between multiple charmonium resonances (as well as potentially open-charm continuum states) are necessary to properly reproduce the pQCD behavior of the $J / \psi \to \pi^0 \gamma^\ast$ TFF in the high-$s$ region. 

\begin{table}[t]
    \centering
    \caption{Same as Table~\ref{tab:Fit} but for the dipole form in Eq.~\eqref{eq:dipole}. We only employ the larger $\left|w_{\psi\phi\pi}\right|$ from Eq.~\eqref{eq:modulus_w}.}
    \label{tab:Fit_v3}
\begin{tabular}{lc}
\hline \hline
Parameters & Theory-Dipole \\
\hline 
$\delta_{\psi\omega\pi}$ & $0.91(27)$ \\
$\delta^{c\bar c}$ & $0.86(20)$ \\
$\Lambda_2~[\rm{GeV}]$ & $5.30(2.39)$ \\
\hline 
$\chi^2 /$ d.o.f. & $27.59/(30-5)=1.02$ \\
\hline \hline
\end{tabular}
\end{table}
\begin{figure}[t]
    \centering
    \includegraphics[width=.5\linewidth]{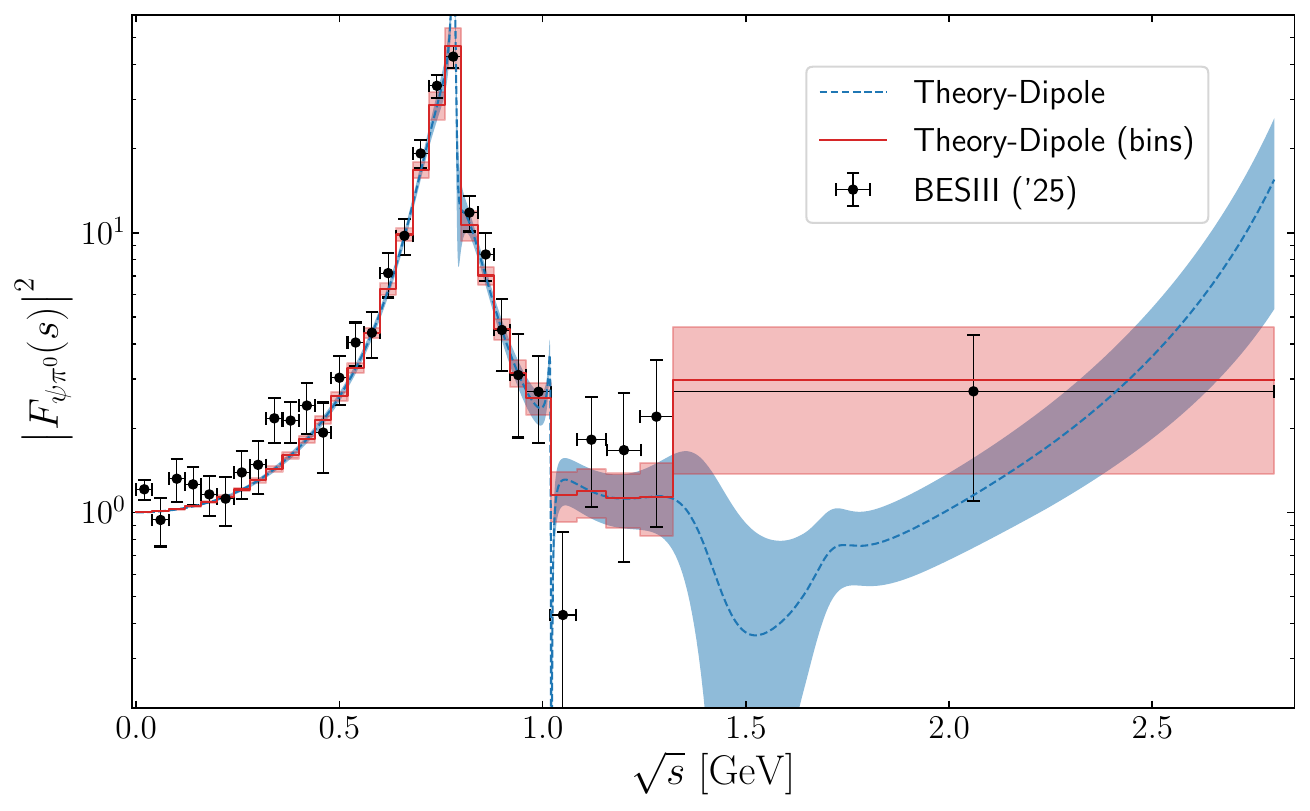}
    \caption{Same as Fig.~\ref{fig:Fit} but for the dipole form in Eq.~\eqref{eq:dipole}.}
    \label{fig:Fit_v3}
\end{figure}
The resulting TFF, including the uncertainties mentioned in the main text, is shown in Fig.~\ref{fig:Fit_v3} and Table~\ref{tab:Fit_v3}.
Comparing with Fig.~\ref{fig:Fit}, one can see that different parameterizations of the charmonium contribution only affect the behavior in the last bin, which corresponds to the region above $\sim 2$~GeV, since the low-energy region is dominated by multipion contributions.

\bibliography{refs}
\end{document}